\documentclass[11pt]{article}
\pdfoutput=1
\usepackage{lipsum}
\usepackage{amsfonts}
\usepackage{graphicx}
\usepackage{epstopdf}
\usepackage{algorithmic}
\ifpdf
  \DeclareGraphicsExtensions{.eps,.pdf,.png,.jpg}
\else
  \DeclareGraphicsExtensions{.eps}
\fi

\title{Unified approach to Floquet lattices, topological insulators, and their nonlinear dynamics
}

\author{Mark Ablowitz\thanks{Department of Applied Mathematics, University of Colorado, Boulder, CO 80301, USA}
\and Justin T. Cole\thanks{Department of Mathematics, University of Colorado, Colorado Springs, CO 80918, USA}  
\and S. D. Nixon\footnotemark[2]
}

\usepackage{amsopn}

\usepackage{mathtools}
\usepackage{amsmath,yhmath}
\usepackage{bm}
\usepackage{graphicx}
\graphicspath{ {./Figures/} }
\usepackage{caption}
\usepackage{subcaption}
\usepackage{soul}
\usepackage{amssymb}

\newcommand{\ii}{\mathrm{i}}
\newcommand{\dd}{\mathrm{d}}
\newcommand{\ee}{\mathrm{e}}
\newcommand{\vv}{\mathrm{\mathbf{v}}}
\newcommand{\rmvec}[1]{\mathrm{\mathbf{#1}}}
\newcommand{\Nu}{N}
\newcommand{\eps}{\varepsilon}
\newcommand{\sign}{\mathrm{sgn}}
\newcommand{\PAvg}[1]{\left\langle #1 \right\rangle}

\newcommand{\uveck}{{\bm{\hat{\textnormal{\bfseries k}}}}}
\newcommand{\disp}{\displaystyle}
\newcommand{\pdv}[2]{\frac{\partial #1}{\partial #2}}

 \usepackage{color}
 \usepackage[dvipsnames]{xcolor}

\newcommand{\sn}[1]{{\color{Purple}$\spadesuit$#1}}

\begin{document}

\maketitle

% REQUIRED
\begin{abstract}
A unified method to analyze the dynamics and  topological structure associated with a class of Floquet topological insulators is presented. The method is applied to a system that describes the propagation of electromagnetic waves through the bulk of a two-dimensional lattice that is helically-driven in the direction of propagation. Tight-binding approximations are employed to derive reduced dynamical systems. Further asymptotic approximations, valid in the high-frequency driving regime, yield a time-averaged system which governs the leading order behavior of the wave. From this follows an analytic calculation of the Berry connection, curvature and Chern number by analyzing the local behavior of the eigenfunctions near the critical points of the spectrum. Examples include honeycomb, Lieb and kagome lattices. In the nonlinear regime novel equations governing  slowly varying wave envelopes are derived. For the honeycomb lattice, numerical simulations show that for relatively small nonlinear effects a striking spiral patterns occurs; as nonlinearity increases localized structures emerge and for somewhat higher nonlinearity the waves collapse.
\end{abstract}

% % REQUIRED
% \begin{keywords}
% topological photonics, chern number, nonlinear, lattice
% \end{keywords}

% % REQUIRED
% \begin{MSCcodes}
% 68Q25, 68R10, 68U05
% \end{MSCcodes}

\section{Introduction}

The theoretical study of topological insulators began nearly 40 years ago with the \textit{Quantum Hall Effect} in electronics \cite{Klitzing1980, Tsui1982}. Topological insulators are materials that behave as an insulator in its interior, but conduct edge waves along their boundaries. An effective  model predicting the same phenomena for periodic systems without any magnetic flux termed the \textit{Quantum Anomalous Hall Effect}), was introduced by Haldane in \cite{Haldane1988}. 
The underlying structure was later reinterpreted to fit a photonic setting where light waves in an optical lattice with suitable permittivity took the place of electrons in magnetic field \cite{Haldane2008}.

The quantization of the Hall conductance can be expressed in terms of invariant integrals \cite{TKNN1982} and gives a unique classification for the geometric structure of the problem \cite{Avron1983}. At the same time, by interpreting the eigenspace of a Hermetian operator as a line bundle over its parameter space \cite{Simon1983}, it follows that these topological
%\maketitle \noindent  
invariants are related to the Berry phase \cite{Berry1984}. In an optical 
setting, this means the Floquet-Bloch spectral bands for a periodic waveguide array, form a natural line bundle over the Brillouin zone. The relevant characterization is the Chern class (and the related Chern number) \cite{Kohmoto1985}.

The broken time-reversal symmetry that characterizes Chern %topological 
insulators and protected edge states has since been realized experimentally using numerous techniques. In \cite{WangChong2008}, the first observation of an electromagnetic protected edge state, capable of propagating around defects, was observed in a magneto-optical setting. In this set up, the symmetry breaking was achieved by the application of an external magnetic field to a square lattice of %magnetized 
ferrite rods.

An experiment in the photonic regime that does not rely on magnetic fields was reported in \cite{Rechtsman2013}.
Here a honeycomb lattice is constructed with waveguides that are helically varying in the direction of beam propagation, which acts as the symmetry breaking mechanism and supports %has a 
topologically protected edge states. 
Under an appropriate change of variables, this has the same effect as a magnetic field in the governing Schr\"{o}dinger equation. 
A tight-binding system in linear and nonlinear regimes was derived in \cite{Curtis2014}.
The field of \textit{Topological Photonics} has since seen many important developments with novel experimental designs that realize spectral bands with underlying nontrivial topological invariants (see review articles \cite{JoannopoulosReview, RechtsmanReview}).

Topological invariants can only change discretely; as a result, induced behavior will be robust against perturbations/defects. 
The phase transition points of the band structure, where the Chern number change, occur at degeneracies of the linear spectrum. 
In a 2D optical lattice, the simplest point degeneracies occur at conically shaped Dirac points in the dispersion bands. Thus, the modulation of honeycomb%or
, Lieb, or kagome lattices, all of which exhibit Dirac/Dirac-like points in their band structure, can lead to nontrivial topologies. The edges states for honeycomb \cite{Curtis2014, Cole2017} and Lieb \cite{Cole2019} lattices with longitudinal driving have been investigated analytically using tight-binding approximations in the lattice nonlinear Schr\"{o}dinger equation (NLS). Here, nonlinear edge solitons are also found to be unidirectional solutions. 
A tight binding model associated with magnetized  ferrite rods was recently derived using Wannier functions in \cite{Cole2020}.

This paper presents a unified approach that starts with the lattice  NLS equations and derives averaged, constant coefficient, models suitable for the analytical study of underlying Floquet-Bloch eigenfunctions,
computation of the Chern number and modeling nonlinear dynamics of the original equation. We use a tight-binding approximation that forgoes the typical Peierls transformation in favor of a more analytically tractable discrete system and further reduce the model by averaging the effect of modulations in the direction of propagation in the bulk (not gap) regime. We apply this method to a wide range of examples including honeycomb, Lieb and kagome lattices in an effort to both illustrate it's ease of use and highlight the commonalities among topological spectra. For honeycomb lattices, it is noteworthy that this averaging technique connects with  the well known Haldane model \cite{Haldane1988} in Brillouin/Fourier space.

In the nonlinear regime, we derive envelope equations for the evolution of wave packets near the degeneracy of the linear spectrum. There are two common spectral phenomena we consider. The first is the typical Dirac point which appears in the spectrum of %for 
the honeycomb and kagome lattices. The second is a Dirac point with a flat band intersecting it as occurs in the Lieb and 1/5 depleted lattices and results in a novel third-order system. In both cases, we investigate the relative effects of driving and nonlinear refraction. 

The nonlinear evolution in the case of a honeycomb lattice is particularly interesting. The equations are a Dirac system with an additional Dirac mass term
which was not present in prior conical diffraction studies \cite{Nixon2009,Zhu2010}. With this extra term numerical computations show that with fixed mass at relatively small nonlinearity there are striking  spiral patterns that gradually disperse; as nonlinearity increases a localized stationary soliton-like structure appears and at yet higher nonlinearity collapse occurs. This similar to what occurs with two dimensional NLS equations cf. \cite{AblowitzRedBook}. 

% CITE: Dirac solitons in optical microresonators

\subsection{Governing Wave Equation}
\label{Sec: Model}

The governing equation for intense paraxial light propagating in an optical lattice is the nonlinear Schr\"{o}dinger equation, which we write in normalized form as
\begin{equation}
   \ii \pdv{\Psi}{z} + \nabla^2\Psi - V(\rmvec{r}, z) \Psi+  \sigma \left|\Psi\right|^2 \Psi= 0
   \label{Eq: Schrodinger}
\end{equation}
where $\Psi$ is the envelope of the electric field, $z$ is the propagation distance, $\rmvec{r}$ is the transverse plane, and the potential $V(\rmvec{r}, z)$ is the index of refraction for a doubly periodic waveguide array in the transverse direction subject to longitudinal driving in the $z$ direction
\begin{equation}
\label{lattice_potential}
   % V(\rmvec{r}, z)= \widetilde{V}(\rmvec{r}-\rmvec{h} (z)).
        V(\rmvec{r}, z)= {V}(\rmvec{r}-\rmvec{h} (z)).
\end{equation}
Optical lattices of this type have been constructed experimentally \cite{Rechtsman2013, Szameit2010} and used in the construction of topological insulators. 
The broken time-symmetry created by rotating the lattice induces an effective magnetic field and yields nontrivial topology in the spectrum and the construction of topologically protected edge modes.

%\sdn{
%\begin{equation}
%\label{lattice_potential}
%   % V(\rmvec{r}, z)= \widetilde{V}(\rmvec{r}-\rmvec{h} (z)).
%        V(\rmvec{r}, z)= {V}(\rmvec{r}-\rmvec{h} (z)).
%\end{equation}
%%
%Optical lattices of this type have been constructed experimentally \cite{Rechtsman2013, Szameit2010} and used in the construction of topological insulators. 
%The broken time-symmetry created by rotating the lattice induces an effective magnetic field and yields nontrivial topology in the spectrum and the construction of protected edge modes.}

The base lattice is doubly periodic in the transverse plane with periods $\vv_1$ and $\vv_2$ and consists of $L$ sublattices. 
A unit cell of the lattice can be pictured as a parallelogram with sides $\vv_1$ and $\vv_2$ and $L$ lattice sites in its interior, one from each sublattice; see Figures \ref{Fig: HCLattice}, \ref{Fig: LiebLattice}, \ref{Fig: KagomeLattice} and \ref{Fig: OneFifthLattice}. 
The full lattice comes from tiling the plane with copies of the unit cell, i.e.,  
\begin{equation}
%   \widetilde{V}(\rmvec{r}) =V_0^2\left[1 ~ - \sum_{(n,m)\in  \mathbb{Z}^2} \sum_{\ell = 1}^L V_L(\bm r - n\, \vv_1 - m\, \vv_2 - \rmvec{d}_\ell)\right]
   V(\rmvec{r}) =V_0^2 ~ - \sum_{(n,m) \in  \mathbb{Z}^2} 
   \sum_{\ell = 1}^L \widetilde{V}(\rmvec{r} - \rmvec{r}_{nm}^\ell)
   \label{Eq: LatticeFull}
\end{equation}
where $\rmvec{r}_{nm}^{\ell} =n~ \vv_1 + m~ \vv_2 + \rmvec{d}_\ell$, the vectors $\rmvec{d}_{\ell}$ define the locations of the sublattice sites within the unit cell and $(n,m)$ is a numbering of the unit cell tiling. For simplicity, we assume that the individual lattice sites have a uniform Gaussian shape,
\begin{equation}
    \widetilde{V}(\rmvec{r}) = V_0^2 \ee^{-\rmvec{r}^2/w}
    \label{Eq: SingleLatticeSite}
\end{equation}
and the lattice sites occur at the minimums of ${V}$. We take driving function of the form
\begin{equation}
    \label{Eq: DrivingFunction}
    \rmvec{h}(z) = \frac{\kappa}{\omega} \Big(\cos(\omega z),~ \sin(\omega z)\Big).
\end{equation}

It is now convenient to move to a frame of reference that follows the lattice, $\widetilde{\rmvec{r}} = \rmvec{r}-\rmvec{h} (z)$, which transforms equation (\ref{Eq: Schrodinger}) into
%c
\begin{equation}
    \ii \pdv{\Psi}{z} -\ii \rmvec{h}'(z) \cdot \widetilde{\nabla} \Psi +  \widetilde{\nabla}^2 \Psi- {V}\big(\widetilde{\rmvec{r}}\big) \Psi +\sigma  \left| \Psi\right|^2\Psi= 0.
   \label{Eq: PSEMovingFrame}  
\end{equation}

In contrast to some previous discrete models for Floquet lattices, cf. \cite{Rechtsman2013, Cole2017, Cole2019}
we do not employ the common Peierls transform. As a useful %an immediate 
consequence, the coefficients of the final averaged system can be expressed in a closed form. For future work, this allows for the study of lattices where the waveguides are not rotating in unison since the $\rmvec{h}'$ term can be treated as the local rotation for the purposes of tight-binding approximations. Results for this non-Peierls approximation agree with those found in models that do take the Peierls transform on lattices that have a moving frame of reference. In Section \ref{Sec: Averaged}, we consider a the fast oscillation, $\omega\gg 1$, regime. Alternate methods instead consider an adiabatic Floquet problem \cite{Sagiv2021}.

\section{Tight-binding approximation}
\label{Sec: TightBinding}

The tight-binding approximation assumes a strong lattice regime where the contrast between the waveguides and the background is large, i.e., $V_0\gg1$ \cite{Zhu2010,Curtis2014, Cole2017}. 
With the lattice sites well-separated, we replace equation (\ref{Eq: Schrodinger}) by a suitable discrete model.
Local behavior near each lattice site is approximated by an exponentially localized orbital function that approximates the discrete eigenmode for a single site lattice times a modulating function that adjusts the local interaction magnitude. 
Due to the rapid decay %localization 
of these orbitals, only the interactions between nearest neighbor lattice sites are considered in the final approximation.

In the neighborhood around $\rmvec{r}_{nm}^{\ell}$, we define $\phi_{nm}^{\ell}(\rmvec{r}) = \phi_0(\rmvec{r} - \rmvec{r}_{nm}^{\ell})$ where $\phi_0(\bm r)$ is the orbital for a lattice site centered at the origin satisfying the equation
\begin{equation}
   \left[ \mu + \nabla^2 + \widetilde{V}(\rmvec{r}) \right] \phi_0(\rmvec{r}) = 0.
\end{equation}
where the orbital function has been normalized to have $L^2(\mathbb{R}^2)$ norm one. The orbital expansion for the wave field, $\psi$, is now given by
\begin{equation}
   \psi(\rmvec{r},z) \approx \sum_{(n,m)\in  \mathbb{Z}^2} \left(\sum_{\ell = 1}^L a_{nm}^{\ell}(z) \phi_{nm}^{\ell}(\rmvec{r}) \right)\ee^{-\ii \left(\mu +V_0^2\right)z}
   \label{Eq: OrbitalExpansion}
\end{equation}
where each orbital function $\phi_{nm}^{\ell}(\rmvec{r})$ is modulated by a related coefficient $a_{nm}^{\ell}(z) $. 
A system of difference equations is obtained by substituting %our 
anatz (\ref{Eq: OrbitalExpansion}) into the modified Schr\"{o}dinger equation (\ref{Eq: Schrodinger}) and taking the inner product of the result with $\phi_{nm}^{\ell}$ for each of lattice sites.

By combining the cluster of lattice sites within each unit cell, $ \rmvec{a}_{nm}(z) = (a_{nm}^{1}(z),\ldots, a_{nm}^{\ell}(z))^T$, we arrive at a nearest neighbor system of difference equations. Note: No interaction occurs more than one unit cell away in any direction. 
\begin{equation}
    \ii \frac{\dd\rmvec{a}_{nm}}{\dd z}  + \sum_{i=-1}^{1}\sum_{j=-1}^{1}\mathcal{L}_{i}^{j}(z)  \rmvec{a}_{n+i,m+j}(z)  + \sigma \rm{N}(\rmvec{a}_{nm})\,\rmvec{a}_{nm}= 0
     \label{Eq: RealSystem}
\end{equation}
where the $\mathcal{L}_i^j$ are a set of connectivity matrices, $\rm{N}(\rmvec{a}) = \rm{diag} \left(|a^{1}|^2,\ldots, |a^{L}|^2\right)$ is the onsite nonlinear response and $\widetilde{\sigma}$ is a constant.
The connectivity matrices are given by the definition
\begin{equation}
    \mathcal{L}_{i}^{j}(z) = \sum_{\ell=1}^{L} \sum_{\rmvec{w}_{\ell k}} \delta_{ij}(\rmvec{w}_{\ell k}) \big[q(z,\rmvec{w}_{\ell k})\big]_{\ell k}
\end{equation}
where the inner sum is taken over all the vectors $\rmvec{w}$ that point from a lattice site on the sublattice $\ell$ to a nearest neighbor on any other sublattice $k$, the delta function $\delta_{ij}(\rmvec{w})$ is $1$ \sn{if the vector $\rmvec{w}$ with starting point $\rmvec{r}_{00}^\ell$ has its end point in the lattice cell $(i, j)$} and $0$ otherwise\sn{,} and $\big[q(z,\rmvec{w})\big]_{\ell k}$ is the matrix with the function $q(z,\rmvec{w})$ in row $\ell$ column $k$ and $0$ everywhere else. For example, the matrix $\mathcal{L}_{1}^{0}(z)$ will have one element, of the form $q(z,\rmvec{w})$, for each lattice site that has a nearest neighbor in the lattice cell one over in the direction $\rmvec{v}_1$, whereas $\mathcal{L}_{0}^{0}(z)$ will have one element for each nearest within the same lattice cell. The connectivity matrices follow the symmetry properties,
\begin{subequations}
\begin{align}
     \mathcal{L}_{i}^{j}(z) &= \Big[\mathcal{L}_{-i}^{~j}(z)\Big]^{H} \\
     \mathcal{L}_{i}^{j}(z) &= \Big[\mathcal{L}_{~i}^{-j}(z)\Big]^{H}
\end{align}
\end{subequations}
where $^{H}$ is the conjugate transpose.
See Appendix \ref{APP: Details} for further details.

The topological properties of the lattice, such as the Chern Number, come from the associated complex vector bundle constructed by taking the reciprocal unit cell, or Brillouin zone, as a base space with attached vector space at each point defined by the eigenspace for a particular band, see Section \ref{Sec: Chern}.  %Thus, 
To carry this out, we take a discrete Fourier transform of the system (\ref{Eq: RealSystem}), i.e, we assume plane waves of the form
\begin{equation}
    \widetilde{\rmvec{a}}_{nm}(z) =  \bm\alpha(\rmvec{k}, z) \ee^{\ii \rmvec{k}\cdot(n \rmvec{v}_1 + m \rmvec{v}_2)},
\end{equation}
where $\rmvec{k}=k_x \hat{\rmvec{i}} + k_y\hat{\rmvec{j}}$ and move the problem to the wavenumber domain. The end result of the tight-binding approximations is a system of nonlinear differential equations, parameterized by $\rmvec{k}$, of the form
\begin{equation}
    \ii \disp \frac{\partial  \bm\alpha}{\partial z } + M\left( z,\rmvec{k} \right) \bm\alpha + \widetilde{\sigma} \rm{N} (\bm\alpha) \bm\alpha= 0.
    \label{Eq: MatrixODE}
\end{equation}
where the matrix $M(z,\rmvec{k})$ is $L \times L$, %$\widetilde{M}(z,\rmvec{k})$ 
%is 
periodic in $z$, doubly periodic in $\rmvec{k}$, Hermitian, due to the symmetry of the pairwise interactions, and has zeros along the diagonal, since any onsite interactions can be removed by a phase transformation. The individual elements of $M(z,\rmvec{k})$ can be written as 
\begin{equation}
    M_{ij}(z,\rmvec{k}) = e^{-\ii \rmvec{k}\cdot(\rmvec{d}_j-\rmvec{d}_i)} \sum_{\rmvec{w}} q(z, \rmvec{w}) e^{\ii \rmvec{k}\cdot\rmvec{w}}
\end{equation}
where
\begin{equation}
    \label{Eq: qz}
    q(z,\rmvec{w}) = q_0 - \ii s_0 \rmvec{h}'(z)\cdot \rmvec{w},
\end{equation}
is a $z$-dependent coefficient function, with $q_0$ and $s_0$ constants, derived in the Appendix \ref{APP: Details} and the sum is taken over the set of vectors, $\rmvec{w}$, that point from a site on sublattice $i$ to a site on sublattice $j$ (see \ref{Fig: HCLattice}, \ref{Fig: LiebLattice}, \ref{Fig: KagomeLattice} or \ref{Fig: OneFifthLattice}).

\section{Averaged System}
\label{Sec: Averaged}
Since equation (\ref{Eq: MatrixODE}) is a system of differential equations with periodic coefficients, it usually does not have an explicit solution. So, we consider the case  where the frequency of the driving $\rmvec{A}(z)$ occur on a faster scale than the background. 
This leads to an averaged system, independent of $z$.
Employing a multiple scales approximation where the lattice matrix is a function of the fast scale: $\zeta =\disp\frac{z}{\varepsilon}$, $0 < \varepsilon \ll 1$,
\begin{equation}
   % M\left( z \right) =  M\left( \disp\frac{z}{\varepsilon}\right),
    M\left( z \right) =  M\left( \disp\zeta \right),
\end{equation}
and the eigenvectors are expanded as $\bm{\alpha} =\bm{\alpha}_0 + \varepsilon \bm{\alpha}_1  + \varepsilon^2 \bm{\alpha}_2\ldots\,$. In terms of the driven potential in Section \ref{Sec: Model}, we define $\omega=\eps^{-1}$. Note: The dependence on $\rmvec{k}$ is not explicitly represented for this analysis, since the steps proceed independently.

Matrix lattice equation (\ref{Eq: MatrixODE}) becomes
\begin{equation}
    \varepsilon^{-1}  \ii \pdv{\bm\alpha}{\zeta}+  \ii \pdv{\bm{\alpha}}{z } + M\left(\zeta \right){\bm\alpha} = 0.
\end{equation}
At $\mathrm{O}(\varepsilon^{-1})$,
\begin{equation}
   \ii  \pdv{\bm\alpha_0}{\zeta} = 0 
\end{equation}
and as a result $\bm\alpha_0$ will be strictly a function of $z$%$\zeta$
, i.e., $\bm\alpha_0(\zeta, z) = \bm\alpha_0(z)$. At $\mathrm{O}(1)$,
\begin{equation}
    \ii\pdv{\bm\alpha_1}{\zeta} + \ii \pdv{\bm\alpha_0}{z}+ M(\zeta) \bm\alpha_0=0.
    \label{Eq: RO Order-1}
\end{equation}
To avoid secular growth in $\bm\alpha_1$, we require that the average effect of the forcing from $\bm\alpha_0$ is at higher order. 
Define $\left< f \right> = \disp\frac{1}{T}\disp\int_0^T f(\zeta) \,\dd \zeta$ where $T$ is the period of $M(\zeta)$. We note that the terms:
$\ii \partial_z \bm{\alpha}_0+\left< M \right>\bm{\alpha}_0$ 
grow without bound in $\zeta$.
Now, the secularity condition can be written as 
\begin{equation}
    \ii \pdv{\bm{\alpha}_0}{z} +  \left<M\right>\bm{\alpha}_0= \eps \rmvec{F}_0.
\end{equation}
Substituting $\ii \partial_z \bm{\alpha}_0 = -\left<M\right>\bm{\alpha}_0$ into equation (\ref{Eq: RO Order-1}) allows us to solve for first correction as
\begin{equation}
    \bm\alpha_1 = \ii M_1(\zeta) \bm{\alpha}_0(z)
        \label{Eq: ROOrder0 2}
\end{equation}
where $M_1(\zeta) = \disp\int_0^\zeta \left( M\left(\widetilde{\zeta}\right) - \left< M \right>  \right) \dd \Tilde{\zeta}$. 

At $\mathrm{O}(\varepsilon)$,
\begin{equation}
    \ii \pdv{\bm{\alpha}_2}{\zeta}+ \ii \pdv{\bm \alpha_1 }{ z} +M(\zeta)\bm{\alpha}_1 + \rmvec{F}_0=0
        \label{Eq: ROOrder1}
\end{equation}
where the terms left over from the previous order must now be balanced as well. 
We can use (\ref{Eq: ROOrder0 2}) to replace $\bm\alpha_1$ and once again substitute $\ii \partial_z \bm{\alpha}_0 = -\PAvg{M}\bm{\alpha}_0$ to simplify the correction terms we will be deriving for the $\bm\alpha_0$ equation. 
These substitutions transform equation (\ref{Eq: ROOrder1}) into 
\begin{align}
  \ii \pdv{\bm\alpha_2}{\zeta}&+ \ii \big( M(\zeta) M_1(\zeta) - M_1(\zeta) \left< M \right> \big) \bm\alpha_0 \nonumber\\
  &+\varepsilon^{-1}\left( \ii \pdv{\bm{\alpha}_0}{z} +  \left<M \right>\bm{\alpha}_0\right)=0 
  \label{Eq: AvgOrder2}
\end{align}
The secularity condition now gives us a final (for our purposes) higher order equation for $\bm\alpha_0$ that will be used to approximate the spectral properties of the lattice,
\begin{equation}
    \ii \pdv{\bm{\alpha}_0}{z} +  \PAvg{M}\bm{\alpha}_0 + \eps \PAvg{\Nu} \bm\alpha_0=0
    \label{Eq: AveragedSystem}
\end{equation}
where $\PAvg{\Nu} = \ii\left[ \left< M M_1\right> - \left< M_1\right>\left< M \right>\right] $. 
The average equation (\ref{Eq: AveragedSystem}) has constant coefficients and hence allows for significant analytic insights; e.g. the computation of the Chern number.

For $M(z,\rmvec{k})$ of the form derived in the tight-binding analysis from Section \ref{Sec: TightBinding} and determined by the nearest neighbor interactions in the lattice, the matrix $\PAvg{\Nu}$ will be determined by all the next-nearest neighbors. This means, the averaged equation at first order is 
\begin{equation}
    \PAvg{M}_{ij} = e^{-\ii \rmvec{k}\cdot(\rmvec{d}_j-\rmvec{d}_i)} \sum_{\rmvec{w}} q_0 e^{\ii \rmvec{k}\cdot\rmvec{w}}
\end{equation}
where $\PAvg{q(z,\rmvec{k})} = q_0$ and  
%$W_{ij}$ are all the vectors pointing from an element of sublattice $i$ to a nearest neighbor on sublattice $j$. 
the elements of the first correct matrix, $\PAvg{\Nu}$, take the form 
\begin{equation}
\label{N_correction}
     \varepsilon \PAvg{\Nu}_{ij} = e^{-\ii \rmvec{k}\cdot(\rmvec{d}_j-\rmvec{d}_i)} \displaystyle\sum_{\rmvec{w}_1,\rmvec{w}_2} Q(\rmvec{w}_1, \rmvec{w}_2) \ee^{\ii \rmvec{k} \cdot (\rmvec{w}_1+\rmvec{w}_2)}
\end{equation}
where the vectors $\rmvec{w}_1$ and $\rmvec{w}_2$ are such that $\rmvec{w}_1$ points from an element of sublattice $i$ to any nearest neighbor and $\rmvec{w}_2$ that starts at the tip of $\rmvec{w}_1$ and ends at an element of the sublattice $j$. So,  the averaged system (\ref{Eq: AveragedSystem}) depends on next-nearest neighbors, even though the original Floquet system (\ref{Eq: MatrixODE}) only depends on nearest neighbor interactions.
The coefficients for the next-nearest neighbor terms are defined by the function 
\begin{align}
   Q(\rmvec{w}_1, \rmvec{w}_2) = \ii \Bigg(&\PAvg{q(z,\rmvec{w}_1) \int_0^z q(\tilde{z},\rmvec{w}_2)-\PAvg{q(\tilde{z},\rmvec{w}_2)}\dd \tilde{z} } \notag\\
   &-\PAvg{\int_0^z q(\tilde{z},\rmvec{w}_1)-\PAvg{q(\tilde{z},\rmvec{w}_1)}\dd \tilde{z}}\PAvg{q(z,\rmvec{w}_2)}\Bigg)
   \notag
\end{align}
which can be found explicitly for the driving function (\ref{Eq: DrivingFunction}) as 
\begin{equation}
    Q(\rmvec{w}_1, \rmvec{w}_2) = \Big( \frac{\ii}{2}s_0^2\kappa^2 (\rmvec{w}_1\wedge\rmvec{w}_2)+q_0s_0\kappa (w_{1x}-w_{2x})\Big).
    \label{Eq: Qwv}
\end{equation}
Here $\wedge$ is the wedge product $(w_x,w_y)\wedge(v_x,v_y) = w_x v_y-v_x w_y$.

\section{Chern Number}
\label{Sec: Chern}

We are now situated to analytically calculate the Chern number from the  related eigenvalue problem $\bm\alpha_0(z,\rmvec{k}) \rightarrow \ee^{-\ii \lambda z} \bm\alpha_0(\rmvec{k})$
%$\bm\alpha_0(z,\rmvec{k}) = \ee^{-\ii \lambda z} \bm\alpha_0(\rmvec{k})$
%%
\begin{equation}
    \lambda \bm{\alpha}_0 + \left(\PAvg{M} + \eps \PAvg{N}\right) \bm\alpha_0 = 0 .
     \label{Eq: AveragedSystemNoTime}
\end{equation}
The number of eigenvalues/eigenvectors depends on the number of sublattices $L$; e.g, for a honeycomb lattice $M$ is a $2\times 2$ matrix.
For a particular band, $\lambda^{\ell}(\rmvec{k})$, the eigenvector, $\bm \alpha_0^{\ell}(\rmvec{k})$, parameterized by the wavenumber, $\rmvec{k}$, forms a natural vector bundle over the torus of the Brillouin Zone \cite{Simon1983}. 
The curvature $\Omega(\rmvec{k})$ of the band (defined below) determines a classification within the de Rham cohomology of the 2D torus \cite{ChernBook}. This is called the Chern class and the Chern number gives an integer indexing of the possible classes.

In concrete terms, for a particular eigenfunction the Berry curvature is given by  
\begin{equation}
    \Omega(\bm k) = \Big( \nabla_{\rmvec{k}} \times
    \bm{\mathcal{A}} \Big) \cdot \widehat{\rmvec{k}}
\end{equation}
where
\begin{equation}
\bm{\mathcal{A}}(\rmvec{k}) = 
\bm \alpha_0^* \cdot \nabla_{\rmvec{k}} \bm\alpha_0,   
\end{equation}
$*$ represents complex conjugate and $\nabla_{\rmvec{k}}=  \hat{\rmvec{i}} \partial_{k_{x}} + \hat{\rmvec{j}} \partial_{k_{y}}$. $\bm{\mathcal{A}}(\rmvec{k})$ is the Berry connection and $\bm \alpha_0$ is chosen such that $\lVert\bm \alpha_0\rVert =1 $. The Chern class of the vector bundle is in fact independent of the choice of connection. For any connection on the vector bundle, the associated curvature is determined from the differential of the connection. While the curvatures vary depending on the choice of connection, they all belong to the same Chern class. It is traditional, however, in the field of \textit{Topological Photonics} to express the curvature in terms of the adiabatic theory developed by Berry \cite{Berry1984}. 

The Chern number is computed from the curvature as 

\begin{equation}
    C = \frac{1}{2\pi\ii}
    \int_{\mathcal{B}}  \Omega(\rmvec{k}) ~\dd k_x \dd k_y
\end{equation}
where $\mathcal{B}$ is the Brillouin zone (reciprocal unit cell).  When the Berry connection is a continuously differentiable function of the wavenumber $\rmvec{k}$ over the entire Brillouin zone, Stokes' theorem may be directly applied to find a zero Chern number, 
\begin{equation}
    C =  \frac{1}{2\pi\ii} \oint_{\partial\mathcal{B}} \bm{\mathcal{A}} \cdot \dd \rmvec{k} = 0,  
    \label{Eq: ChernConnection}
\end{equation}
since the connection, $\bm{\mathcal{A}}$, has period matching that of $\mathcal{B}$. 
This means nonzero Chern numbers are associated with discontinuities in the connection which in turn are associated with discontinuities which appear in the phase of of the eigenfunction $\bm \alpha_0$. 

Global information about the curvature is found from local behavior of the Berry connection by isolating and removing discontinuities of the spectrum from the region of integration, $\mathcal{B}$. 
This can be done without effecting the Chern number since the curvature is bounded and vanishing small regions in the domain of integration correspond to vanishing small contributions to the integral. 
The Berry curvature is gauge invariant, however, this not true of the connection where a transformation in the eigenfunction $\bm \alpha_0(\rmvec{k}) \rightarrow \ee^{\ii \chi(\rmvec{k})}\bm \alpha_0(\rmvec{k})$ results in a corresponding transform in the connection, $\mathcal{A} \rightarrow \mathcal{A} + i \nabla_{\bm k} \chi(\bm k)$. 
As a consequence, the discontinuities in the connection may be arbitrarily positioned within the Brillouin zone. To remove this ambiguity the eigenfunction can be chosen to have a constant phase in one component. 
The location of discontinuities will now be uniquely defined and may be dealt with analytically. 

Alternatively, the Chern number may be obtained numerically using the method presented in \cite{Hatsugai05}. This method similarly relies on a fixed choice of gauge, however it does not require knowledge of where the discontinuities of the Berry phase lie. Thus, for higher order systems where a closed form for the eigenvectors can't be easily obtained, a numerical calculation for the Chern number will suffice. Nonetheless, we note that in principal the calculation of the Chern number can be carried out analytically as demonstrated in Section \ref{Sec: Lieb}.

\subsection{Honeycomb Lattice}
\label{Sec: Honeycomb}

\begin{figure}[ht]
    \centering
            \includegraphics[width=0.55\textwidth]{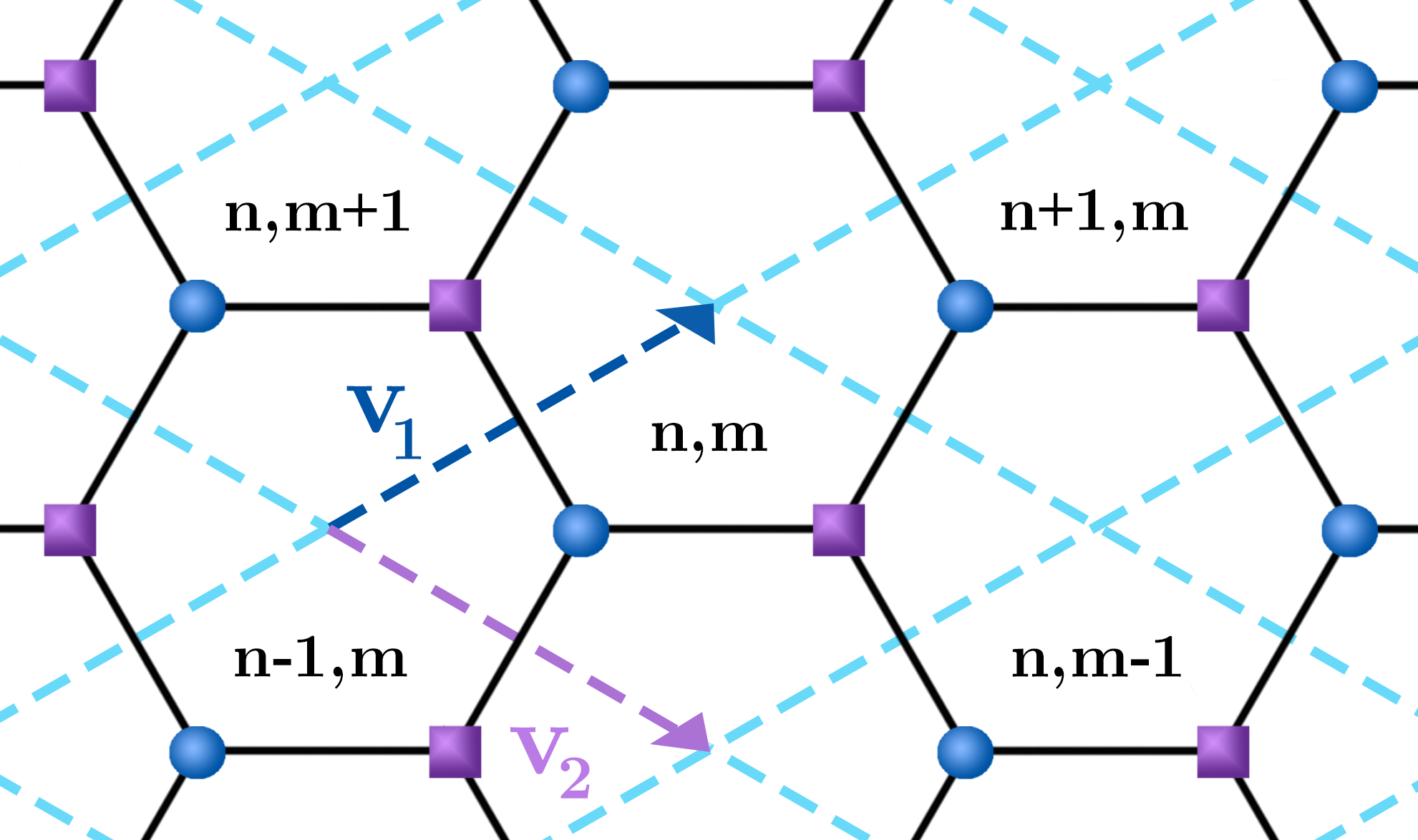}
            \hspace{1cm}
            \includegraphics[width=0.2\textwidth]{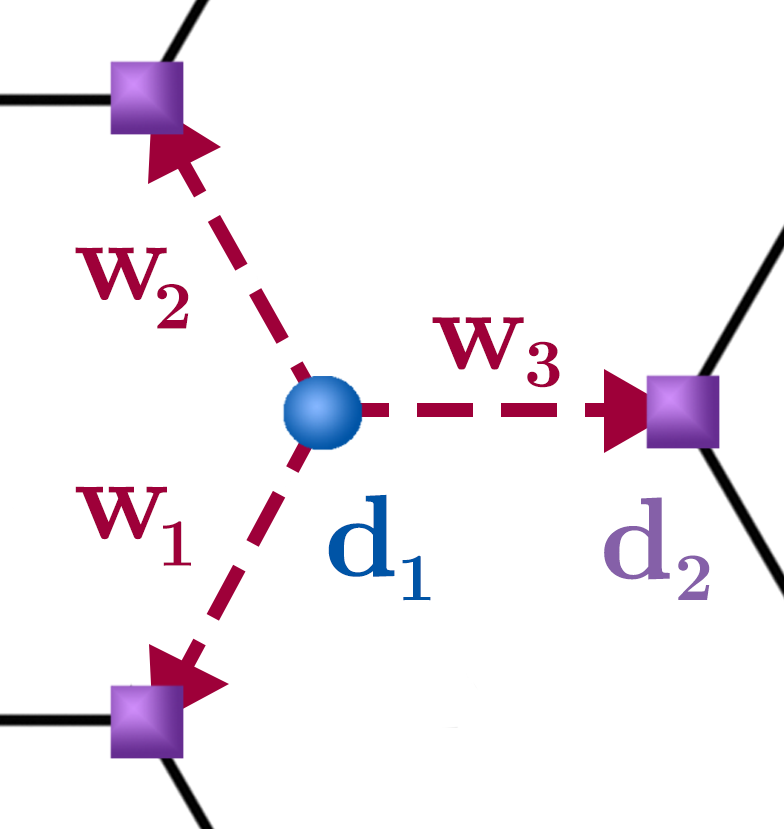}
            \caption{The honeycomb lattice consists of two triangular sublattices. Blue spheres represent the first sublattice and purple squares represent the second sublattice. Dotted lines outline the lattice cells.}
            \label{Fig: HCLattice}
\end{figure}

The honeycomb lattice, named for its hexagonal tiling, is well known for its connections to graphene \cite{GrapheneReview} and for being the simplest lattice model giving rise to Dirac points \cite{Weinstein}.
It can be thought of as two interlaced triangular sublattices. (See Figure \ref{Fig: HCLattice}). 
For the honeycomb lattice, there are two Dirac points in the Brillouin zone.
Thus, there are two possible locations for discontinuities to naturally occur in the eigenvectors. Although this doesn't effect the Chern number, the expression for the eigenvectors does have consequences for the Berry connection and the local behavior being used to arrive at the Chern number analytically.

\begin{figure}[ht]
    \centering
        \begin{subfigure}[b]{0.38\textwidth}
            \centering
            \includegraphics[width=0.9\textwidth]{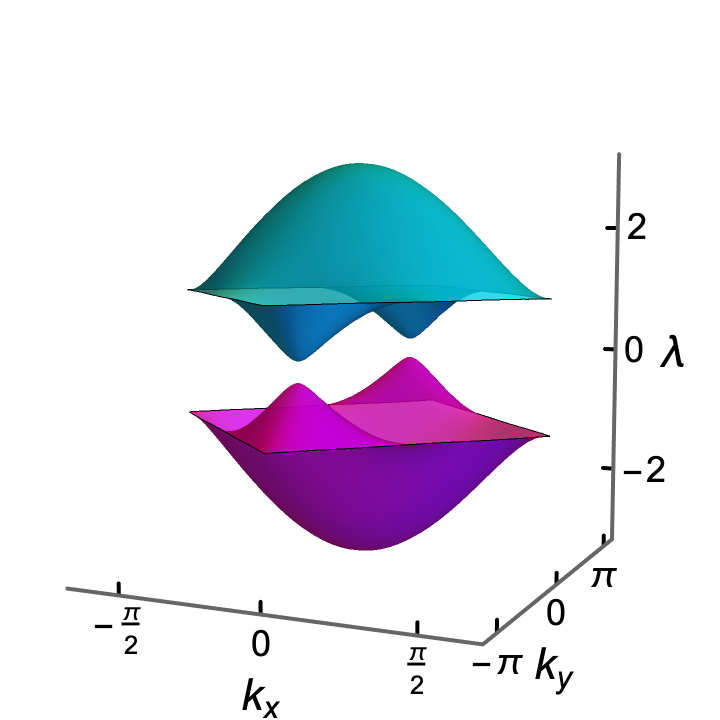}
            \caption{Spectrum for the honeycomb lattice with $\rho = 1$.
            }
        \end{subfigure}
        \hspace{0.4cm}
        \begin{subfigure}[b]{0.38\textwidth}
            \centering
            \includegraphics[width=0.9\textwidth]{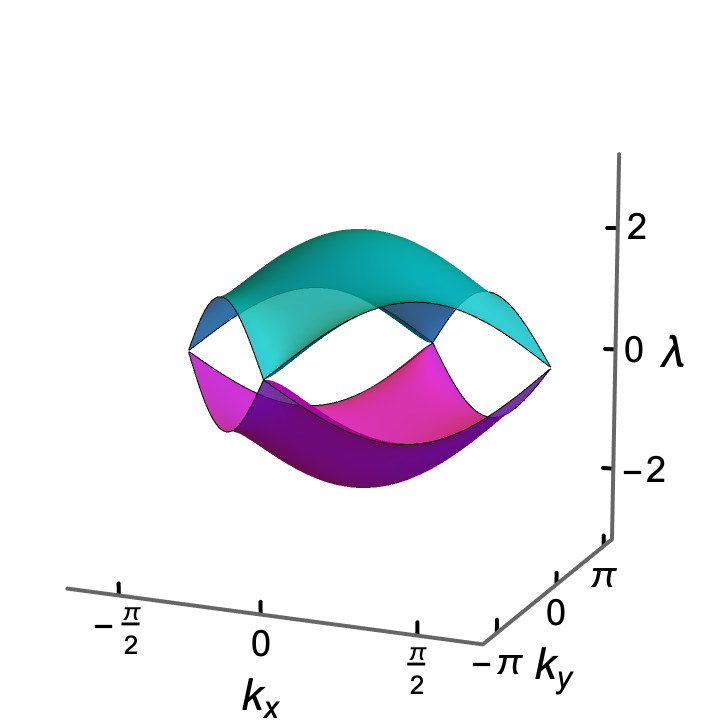}
            \caption{Spectrum for the honeycomb lattice with $\rho = 0.5$.}
             \label{Fig: HCSpectrum2}
        \end{subfigure}
    \caption{Bulk spectral surfaces %structure 
    for the honeycomb lattice over one Brillouin zone.}
    \label{Fig: HCSpectrum}
\end{figure}

In terms of the framework of Section \ref{Sec: Model}, the Honeycomb lattice is defined by $\vv_1 = \left(\frac{3}{2},~ \frac{\sqrt{3}}{2}\right)$, $\vv_2 = \left(\frac{3}{2}, -\frac{\sqrt{3}}{2}\right)$ and $\mathrm{\mathbf{d}}_1 = (1,0)$, $\mathrm{\mathbf{d}}_2 = (2,0)$ and $\rmvec{w}_1 = \left(-\frac{1}{2}, -\frac{\sqrt{3}}{2}\right)$, $\rmvec{w}_2 = \left(-\frac{1}{2}, \frac{\sqrt{3}}{2}\right)$, $\rmvec{w}_3 = \left(1,0\right)$, where the distance between nearest neighbors is taken to be one. In the wavenumber domain of equation (\ref{Eq: MatrixODE}), the system for our honeycomb lattice is 

\begin{equation}
    \ii\frac{\dd \bm{\alpha}}{\dd z} + 
        \begin{bmatrix}
            0 & \gamma^*(z,\rmvec{k}) \\
            \gamma(z,\rmvec{k})  & 0
        \end{bmatrix}
    \bm\alpha =0
    \label{Eq: HoneycombFull}
\end{equation}
where 
\begin{equation}
\gamma(z,\rmvec{k}) = q(z,\rmvec{w_3}) + \rho \Big( q(z,\rmvec{w_1}) \ee^{\ii \vv_1 \cdot \rmvec{k}}+q(z,\rmvec{w_2}) \ee^{\ii \vv_2 \cdot \rmvec{k}} \Big)
\end{equation}
and $q(z,\rmvec{w})$ is the coefficient function (\ref{Eq: qz}) defined earlier. The parameter $\rho$ represents a stretching/contracting of the lattice in the $y$-direction, in which case the strength of the connection along the vectors $\rmvec{w}_1$ and $\rmvec{w}_2$ differs from the strength of the connection along the vector $\rmvec{w}_3$. This amounts to a breaking of the $120^{\circ}$ rotational symmetry of the  lattice.

Substituting the matrix from (\ref{Eq: HoneycombFull}) into the general averaged system (\ref{Eq: AveragedSystem}), we have the averaged matrix
\begin{equation*}
    \PAvg{M } = 
        \begin{bmatrix}
            0 & \langle\gamma\rangle^*(\rmvec{k}) \\
            \langle\gamma\rangle(\rmvec{k})  & 0
        \end{bmatrix}
\end{equation*}
where 
\begin{equation}
    \PAvg{\gamma}(\rmvec{k}) = q_0 \left(1 + \rho ( \ee^{\ii \rmvec{v}_1 \cdot \rmvec{k}} +  \ee^{\ii \rmvec{v}_2 \cdot \rmvec{k}})  \right)
    \label{Eq: AvgMHoneycomb}
\end{equation}
and the constant $q_0 = \PAvg{q(z,\rmvec{w}})$ for all $\rmvec{w}$ is the same as defined in Section \ref{Sec: TightBinding} and can be factored out.
The order $\eps$ correction term is given by
\begin{equation}
    \PAvg{\Nu}(\rmvec{k}) =
        \begin{bmatrix}
        \nu(\rmvec{k}) & 0 \\
        0 & -\nu(\rmvec{k}) 
        \end{bmatrix}
        \label{Eq: AvgNHoneycomb}
\end{equation}
where
\begin{align}
 \nu(\rmvec{k}) = \rho  Q(\rmvec{w}_1,-\rmvec{w}_3)&\ee^{\ii \rmvec{k} \cdot \rmvec{v}_1 } + \rho^2 Q(\rmvec{w}_2,-\rmvec{w}_1)\ee^{\ii \rmvec{k} \cdot (\rmvec{v}_2 -\rmvec{v}_1)}\\
&+ \rho Q(\rmvec{w}_3,-\rmvec{w}_2)\ee^{-\ii \rmvec{k} \cdot \rmvec{v}_2} + c.c. \notag
\end{align}
and ``c.c." stands for the complex conjugate, so $\nu(\rmvec{k})$ is a real valued function.
Combining equations (\ref{Eq: AvgMHoneycomb}) and (\ref{Eq: AvgNHoneycomb}) we arrive at the eigenvalue problem of the averaged system from equation (\ref{Eq: AveragedSystemNoTime})
\begin{equation}
   \lambda \bm{\alpha}_0 + 
        \begin{bmatrix}
            \eps \nu(\rmvec{k}) & \PAvg{\gamma}^*(\rmvec{k}) \\
            \PAvg{\gamma}(\rmvec{k})  & -\eps \nu(\rmvec{k}) 
        \end{bmatrix}
    \bm\alpha_0 =0.
    \label{Eq: HoneycombAveraged}
\end{equation}
with eigenvalues
\begin{equation}
    \lambda(\rmvec{k}) = \pm\sqrt{\lvert  \PAvg{\gamma}\rvert^2 + \eps^2 \lvert\nu\rvert^2}
\end{equation}
This averaged equation for the honeycomb lattice is equivalent to the well-known Haldane model \cite{Haldane1988} in Fourier space; this is due to the fact that  
the first correction in the averaging introduces the next-nearest neighbor terms that are used to break time reversal symmetry in Haldane's model. In terms of the Pauli matrices, equation (\ref{Eq: HoneycombAveraged}) can be written as 
\begin{align}
 \lambda I \bm{\alpha}& + \sigma_x q_0\Big(1 + \rho \cos(\rmvec{k}\cdot\rmvec{v}_1) + \rho \cos(\rmvec{k}\cdot\rmvec{v}_2)\Big) \\ &+ \sigma_y q_0 \Big(1 + \rho \sin(\rmvec{k}\cdot\rmvec{v}_1) + \rho \sin(\rmvec{k}\cdot\rmvec{v}_2) \Big)\notag \\
&-\frac{1}{\omega} \sigma_z  s_0^2 \kappa^2 \frac{\sqrt{3}}{2} \Big( \rho \sin(\rmvec{k}\cdot \rmvec{v}_1) +  \rho^2 \sin(\rmvec{k}\cdot (\rmvec{v}_2-\rmvec{v}_1)) + \rho \sin(-\rmvec{k}\cdot \rmvec{v}_2 ) \Big)\notag\\
&+ \frac{1}{\omega} \sigma_z q_0 s_0 \kappa \Big(  \rho \cos(\rmvec{k}\cdot \rmvec{v}_1) -  2\rho^2 \cos(\rmvec{k}\cdot (\rmvec{v}_2-\rmvec{v}_1)) + \rho \cos(-\rmvec{k}\cdot \rmvec{v}_2) \Big)\notag = 0.
\end{align}
This corresponds to the Haldane model with zero mass, i.e. inversion symmetry. 

Below we give the analytic step for calculating the Chern number within the framework of Section \ref{Sec: TightBinding} and this paper in general. The eigenvectors of equation (\ref{Eq: HoneycombAveraged}) are
\begin{equation}
\label{efHC}
    \bm \alpha_0 = \frac{1}{D(\rmvec{k})}
        \begin{bmatrix}
            \PAvg{\gamma}^*\\
            -\left(\lambda +\eps\nu \right)
        \end{bmatrix}
\end{equation}
where $D(\rmvec{k})$ is a normalizing term such that $ \bm\alpha_0^{*} \cdot \bm\alpha_0 = 1$; it is given by
\begin{align}
    D(\rmvec{k})^2   & = \left|\PAvg{\gamma}\right|^2 + \left(\lambda +\eps \nu \right)^2\notag\\
                & = \left|\PAvg{\gamma}\right|^2 + \lambda^2 + 2 \eps \lambda \nu + \eps^2 \nu \notag\\
                & = 2 \lambda \left(\lambda + \eps\nu \right)
\end{align}
Note: As a result of the longitudinal driving ($\varepsilon \not= 0 $), the honeycomb spectrum in equation (\ref{Eq: HoneycombAveraged}) is no longer conical at the Dirac point. Furthermore, at the Dirac point, where $\PAvg{\gamma}(\rmvec{K}_D) = 0 $, there is a gap width of $ 2 \lambda(\rmvec{K}_D) = 2 \varepsilon \nu$.

\begin{figure}[ht]
    \centering
        \begin{subfigure}[b]{0.4\textwidth}
            \centering
            \includegraphics[width=\textwidth]{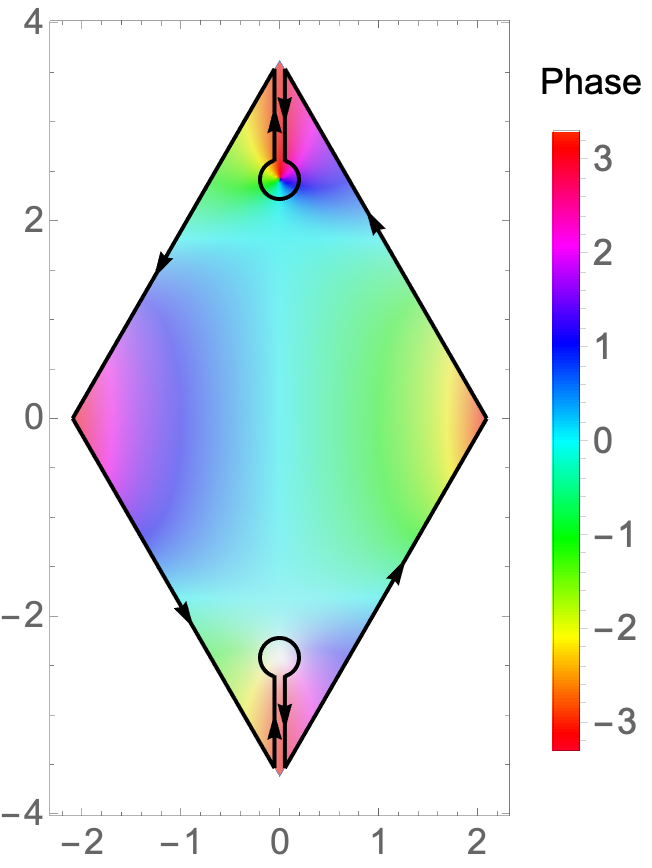}
            \caption{Magnitude and phase for\\ the first
            component of ${\bm \alpha}_0$.}
            \label{Fig: HCAlpha1Top}
        \end{subfigure}
        \hspace{0.2cm}
        \begin{subfigure}[b]{0.33\textwidth}
            \center
            \includegraphics[width=0.955\textwidth]{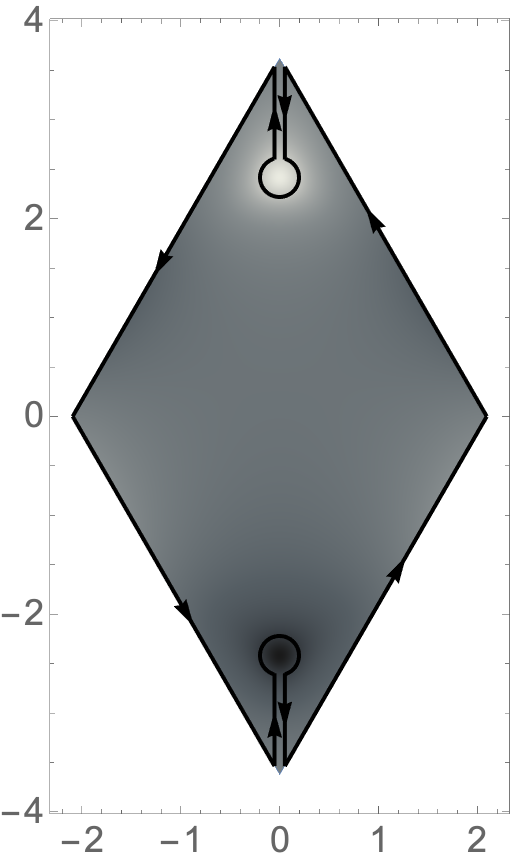}
            \caption{Magnitude for the second component of ${\bm \alpha}_0$.}
            \label{Fig: HCAlpha2Top}
        \end{subfigure}
    \caption{Contour of integration that removes the critical points at $\rmvec{K}_D = \left(0,\frac{4\pi}{3\sqrt{3}}\right)$ and $\rmvec{K}'_D = \left(0,-\frac{4\pi}{3\sqrt{3}}\right)$. The magnitudes and phases are shown for the components of $\bm\alpha_0$ (\ref{efHC}) on the upper band. The magnitudes are represented by color saturation (with white representing a magnitude of $0$) and the phases are represented by hue (with the second component being strictly real and shown in grey scale).} 
    \label{Fig: HCContour}
\end{figure}

At any point where $\PAvg{\gamma}(\rmvec{K}_D) = 0 $, and therefore $\lambda = \pm \eps \left| \nu \right|$, the vanishing of the normalizing function leads to  a discontinuity in the eigenvector. 
So, at a critical point $\rmvec{K}_D$,
\begin{equation}
    D(\rmvec{K}_D)^2 = 2 (\eps \nu)^2\left( 1 + \sign(\lambda \nu )\right)
\end{equation}
and thus $\bm \alpha_0$ has an integrable singularity if $\sign(\nu) = -\sign(\lambda)$. 

As an example, we take $\rho = 1$. The critical points now occur at $\rmvec{K}_D = \left( 0,\,\frac{4 \pi}{\sqrt{3}} \right)$ and $\rmvec{K}'_D = \left( 0,-\frac{4 \pi}{\sqrt{3} }\right)$. We look for the local behavior of the eigenvectors near one of these points, as well as the Berry connection and curvature which are derived from them. For illustrative proposes, we take the expansion near
\begin{equation}
\rmvec{k} = \rmvec{K}_D + \frac{ 1}{q_0}\delta \rmvec{k}    
\end{equation} 
where $\left\lVert\delta \rmvec{k}\right\rVert \ll 1$ and the expansion around $\bm K_D'$ follows similarly. 
Approximations for the key terms are 
\begin{subequations}
\begin{align}
\PAvg{\gamma}(\rmvec{k}) &\approx -\frac{3}{2}(\ii \delta k_x + \delta k_y) \\
 \nu(\rmvec{k}) &\approx \nu(\rmvec{K}_D) \equiv \nu_0\\
\lambda(\rmvec{k}) &\approx \pm \sqrt{\frac{9}{4}(\delta k_x^2 + \delta k_y^2)+ \eps^2\nu_0^2} 
\end{align}
\end{subequations}
where $\nu_0 \in \mathbb{R}$. Substituting these approximations into $\bm \alpha_0$ gives 
\begin{equation}
    \bm \alpha_0(\rmvec{k}) \approx -\frac{1}{\sqrt{2 \lambda}} 
        \begin{bmatrix}
            \disp\frac{\frac{3}{2}(\delta k_y - \ii\delta k_x)}
            {\sqrt{\lambda + \eps \nu_0}}\\
            \sqrt{\lambda + \eps \nu_0}
        \end{bmatrix}.
\end{equation}

The local approximation of the Berry connection is
\begin{equation}
\bm{\mathcal{A}}(\rmvec{k}) \approx 
\disp\frac{9 \ii }{8 \lambda (\lambda + \eps \nu_0)}(-\delta k_y \hat{\rmvec{i}} + \delta k_x \hat{\rmvec{j}}).
\end{equation}
This shows that the discontinuity from the eigenvector carries through to the connection in the form of a singularity that behaves like $\frac{1}{\lVert \delta \rmvec{k} \rVert}$ when $\mathrm{sgn}(\lambda \nu_0) = -1$. By a similar set of calculation, we find that a potential singularity exists at $\rmvec{K}'_D$ as well. 
The approximation of the Berry curvature is
\begin{equation}
    \Omega(\rmvec{k}) \approx 
    - \eps\disp\frac{9\nu_0\ii}{8\lambda^3}\uveck .
\end{equation}
Note that $|\lambda| >0$ and so the curvature is bounded. Due to the radial symmetry of $\bm{\mathcal{A}}\cdot\, \dd \rmvec{k} $ around the critical points, it's convenient to define the disk regions
\begin{equation}
    \mathcal{D}(\bm \delta) = \left\{ \bm k \middle| \left\lVert \bm K_D- \bm k\right\rVert \leq \delta \right\} , \mathcal{D}'(\bm \delta) = \left\{ \bm k \middle| \left\lVert \bm K'_D- \bm k\right\rVert \leq \delta \right\} 
\end{equation}
with the punctured Brillouin zone
\begin{equation}
    \iint_{\mathcal{B}} = \disp\lim_{\delta \rightarrow 0} \iint_{\mathcal{B} -\mathcal{D}(\bm \delta) - \mathcal{D}'(\bm \delta)}.
\end{equation}
In this way, we may carve out a small region around any potential singularities without effecting the the calculation of the curvature or the Chern number. Once all singularities have been excised from the Brillouin zone, by removing infinitesimal regions around $\rmvec{K}_D$ and $\rmvec{K}'_D$, we can apply Stokes theorem to represent global information about the curvature in terms of the local behavior of connection near the critical points 
\begin{equation}
    \iint_{\mathcal{B}} \Omega(\rmvec{k}) \cdot \uveck   \, \dd k_x \dd k_y=\oint_{\partial\mathcal{B}} \mathcal{A}(\rmvec{k})  \, \dd \rmvec{k} -C_{\rmvec{K}_D}-C_{\rmvec{K}_D'} 
\end{equation}
where
\begin{subequations}
\begin{align}
    C_{\rmvec{K}_D} & = \lim_{\delta\rightarrow 0}  \oint_{\partial\mathcal{D}(\delta)} \mathcal{A}(\rmvec{k})  \, \dd \rmvec{k}\\
    C_{\rmvec{K}'_D}& = \lim_{\delta\rightarrow 0}  \oint_{\partial\mathcal{D}'(\delta)} \mathcal{A}(\rmvec{k})  \, \dd \rmvec{k} .
\end{align}
\end{subequations}
An illustration of the contours taken can be found in figure \ref{Fig: HCContour}. Here we take $
\bm \alpha_0$ on the upper band. At $\rmvec{K}_D$ the first component of the eigenvector approaches a magnitude of 1 and exhibits a $2\pi$ change in phase around this point. This is the discontinuity that must be integrated around. By contrast, the first component of the eigenvector is zero at $\rmvec{K}'_D$ and same $2\pi$ phase change no longer corresponds to a discontinuity. On the lower band this is reversed, which is why we have chosen to remove both points from the domain. From the periodicity of the eigenvectors in $\rmvec{k}$, the boundary integral around $\partial\mathcal{B}$ is zero.

Define $\nu_0 = \nu(\rmvec{K}_D)= -\nu(\rmvec{K}'_D)$. The Chern number is found to be
\begin{align*}
    C  =& \frac{1}{2\pi \ii}\left(-C_{\rmvec{K}_D}- C_{\rmvec{K}'_D}\right)\\
     =& \disp\lim_{\bm {\delta}\rightarrow 0} \Bigg[-\left(\frac{1}{2} - \sign(\lambda) \disp\frac{\eps \nu_0}{\sqrt{9 \delta^2 + 4 (\eps \nu_0)^2}}\right)\notag\\
     &\quad +\left(\frac{1}{2} + \sign(\lambda) \disp\frac{\eps \nu_0}{\sqrt{9 \delta^2 + 4 (\eps \nu_0)^2}}\right)\Bigg]\\
     =&\, \sign(\lambda \nu_0).
\end{align*}

For $\rho = 1$, the Chern number is $-1$ for the upper band, $\lambda>0$, and $+1$ for the lower band, $\lambda < 0$. This is consistent with results from \cite{Cole2019, ChernAnalytic}.
This topology persists for $\rho > 1/2$. When $\rho = 1/2$ the two critical points merge at the corner of the Brillouin zone, i.e., $\rmvec{K}_D = \rmvec{K}'_D = \left(0, \pm \frac{4\pi}{\sqrt{3}}\right)$, as seen in Figure (\ref{Fig: HCSpectrum2}). At this point $\nu(\rmvec{K}_D )= 0$ and so the gap in the spectrum caused by the driving of the lattice closes and a Dirac point forms. For $\rho<1/2$, a gap opens that is caused by the stretching of the lattice instead of the driving of the lattice. In the analysis, this corresponds to $|\gamma|>0$ for all $\rmvec{k}$ in the Brillouin zone, so that no discontinuities exist in the phase of the eigenvectors and the Chern Number on both bands is zero.

\subsection{Lieb Lattice}
\label{Sec: Lieb}

\begin{figure}[ht]
    \centering
        \begin{subfigure}[b]{0.52\textwidth}
            \centering
            \includegraphics[width=0.6\textwidth]{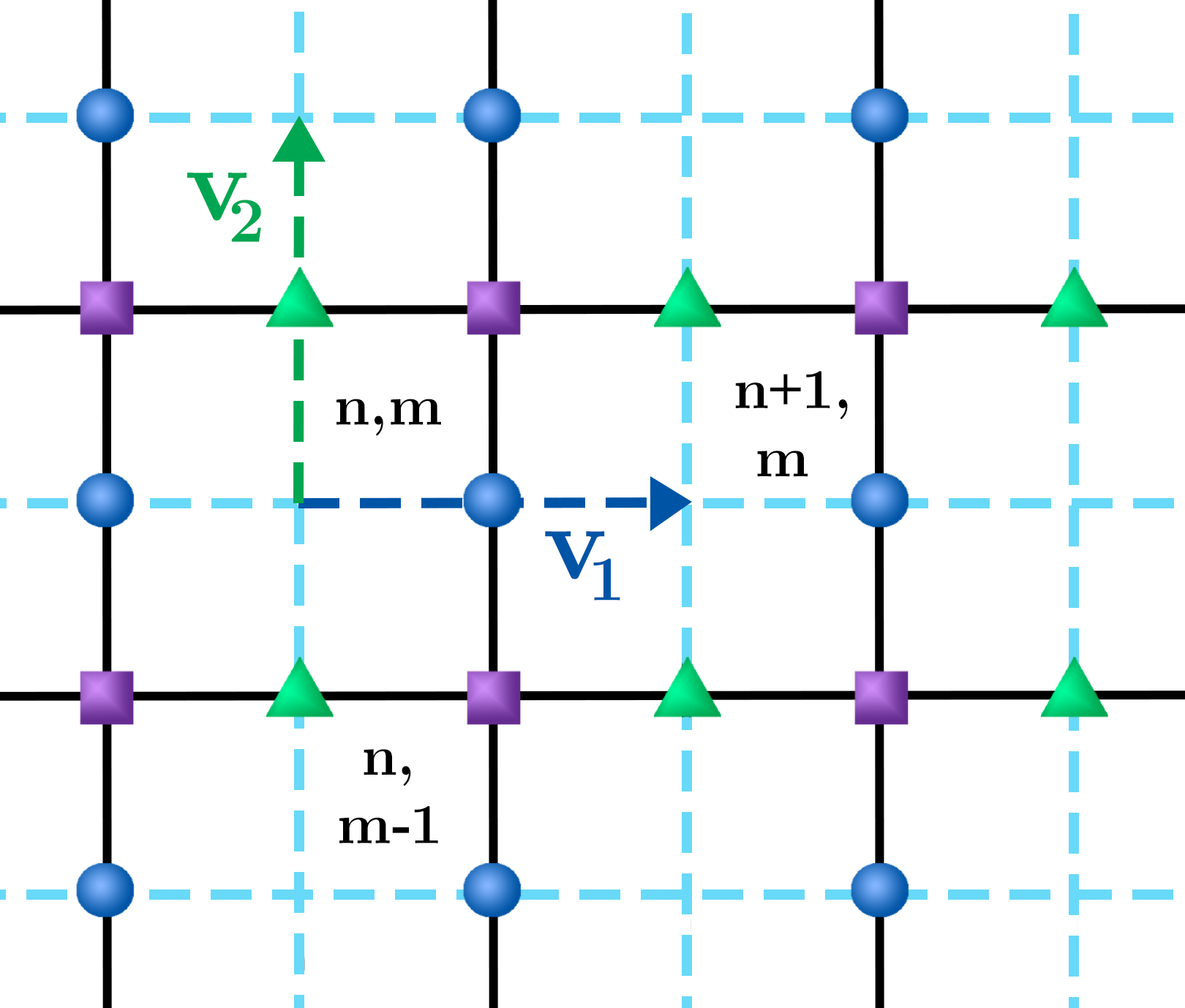}
            \hspace{0.1cm}
            \includegraphics[width=0.35\textwidth]{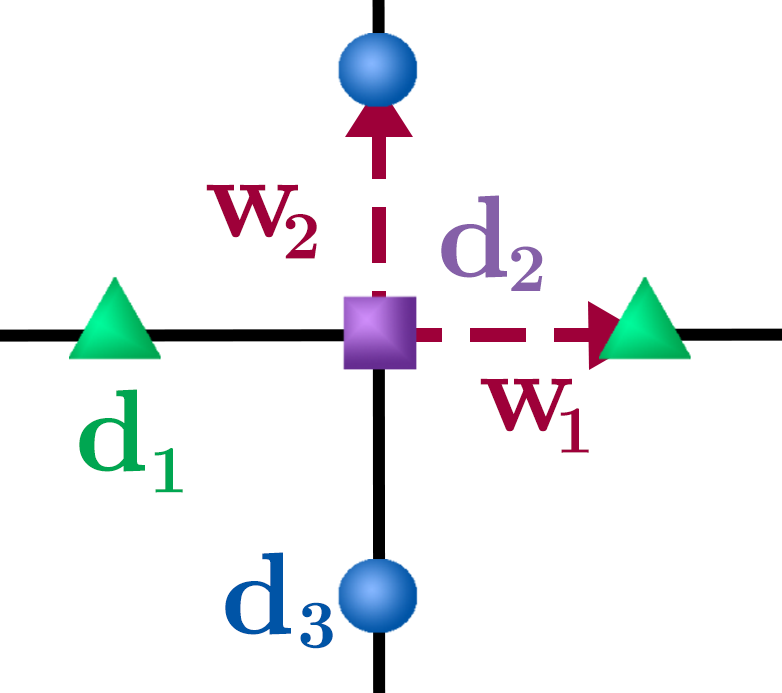}
            \caption{Diagram for the Lieb lattice with three lattice sites per unit cell.}
            %\vspace{0.3cm}
            \label{Fig: LiebLattice}
        \end{subfigure}
        \hspace{0.5cm}
        \begin{subfigure}[b]{0.32\textwidth}
            \centering
            \includegraphics[width=0.9\textwidth]{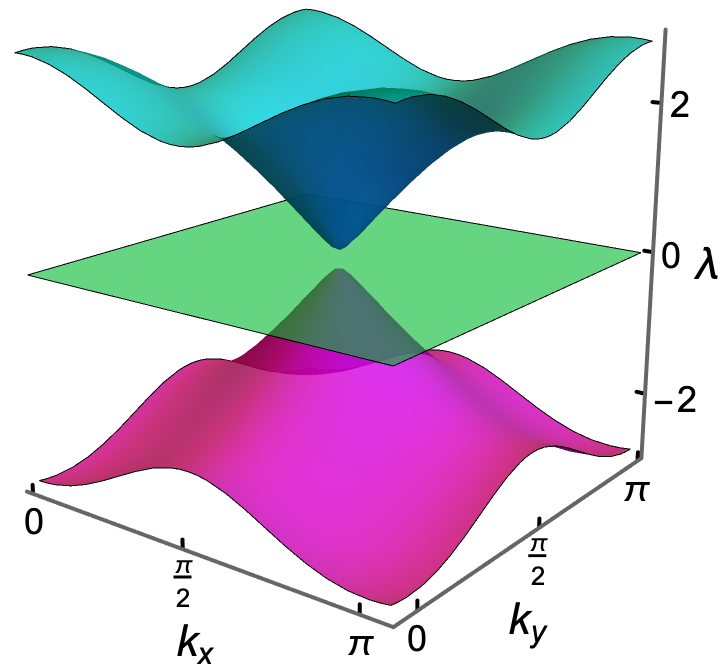}
            \caption{Spectrum for the Lieb lattice.
            }
            \vspace{0.3cm}
        \end{subfigure}
    \caption{The Lieb Lattice consists of three square sublattices. The green triangles represent the first sublattice, the purple squares 
    represent the second sublattice and the teal spheres represent the third sublattice. }
    \label{Fig: LiebSpectrum}
\end{figure}

Here we consider the Lieb lattice \cite{Lieb1989,Franz2010} as an example. The Lieb lattice is a 1/4-depleted square lattice, i.e., one in four lattice sites have been removed, consisting of three sublattice sites in each unit cell. The Lieb system features a single spectral degeneracy within the Brillouin zone where upper and lower bands intersect with a middle flat-band at a Dirac point (see Figure \ref{Fig: LiebSpectrum}(b)). While the three distinct sub-lattices result in a 3rd order system in %the 
$\rmvec{k}$-space, the lone Dirac point streamlines the calculations of the Chern number. As we will show, the Dirac point lines up with a singularity in the Berry connection that occurs on both the upper and lower bands. 

In terms of the framework laid out in Section \ref{Sec: Model}, the Lieb lattice is defined by $\vv_1 = (2,0)$, $\vv_2 = (0,2)$ and $\rmvec{d}_1 = (0,1)$, $\rmvec{d}_2 = (1,1)$, $\rmvec{d}_3 = (1,0)$ and $\rmvec{w}_1 = (1,0)$, $\rmvec{w}_2  = (0, 1)$. The system of differential equations, in the form (\ref{Eq: MatrixODE}), derived for this lattice is
\begin{equation}
    \ii \frac{\dd \bm{\alpha}}{\dd z}+ 
        \begin{bmatrix}
            0 & \gamma_1^*(z,\rmvec{k})& 0 \\
            \gamma_1(z,\rmvec{k}) & 0 &      
                \gamma_2^*(z,\rmvec{k})\\
            0 & \gamma_2(z,\rmvec{k})& 0 
        \end{bmatrix}
    \bm \alpha = 0
    \label{Eq: LiebFull}
\end{equation}
where the $\gamma$ functions are given by 
\begin{subequations}
\begin{align}
\gamma_1(z , \rmvec{k})  &=  q(z, \rmvec{w}_1) \ee^{\ii 2 k_x}+ q(z, -\rmvec{w}_1)\\
\gamma_2(z , \rmvec{k})  &=  q(z, \rmvec{w}_2) \ee^{\ii 2 k_y}+ q(z, -\rmvec{w}_2)
\end{align}
\end{subequations}
and $q$ as defined in equation (\ref{Eq: qz}).

Looking at Figure \ref{Fig: LiebSpectrum}, we see that there is no interaction between the first and third sublattice, which correspond to lattice sites positioned between the lattice sites of the square lattice created by the second sublattice. This lack of interaction is reflected in the zeros found in the upper right and lower left corners of the matrix from equation (\ref{Eq: LiebFull}). A detailed derivation of the specific $\gamma$ functions given here as well as a more generalized version of this system appears in \cite{Cole2019}.

Substituting the matrix from the Lieb system (\ref{Eq: LiebFull}) into the general averaged equation (\ref{Eq: AveragedSystem}) gives the averaged matrix
\begin{equation}
    \PAvg{M}(\rmvec{k})=
        \begin{bmatrix}
            0 & \PAvg{\gamma_1}^*(\rmvec{k})& 0 \\
            \PAvg{\gamma_1}(\rmvec{k}) & 0 &      
                \PAvg{\gamma_2}^*(\rmvec{k})\\
            0 & \PAvg{\gamma_2}(\rmvec{k})& 0 
        \end{bmatrix}.
\end{equation}
For the averaged $\gamma_i$ functions, we again use the fact that $ \PAvg{q(\zeta, \rmvec{w})} = q_0$. This gives 
\begin{subequations}
\begin{align}
    \PAvg{\gamma_1}(\rmvec{k})  &= q_0 \left( \ee^{\ii 2 k_x}+1\right)\\
    \PAvg{\gamma_2}(\rmvec{k})  &= q_0 \left( \ee^{\ii 2 k_y}+1\right)
\end{align}
\end{subequations}
At order $\eps$, the correction term is
\begin{equation}
\label{Lieb_N_correction}
\PAvg{\Nu}(\rmvec{k})  = 
    \begin{bmatrix}
        0 & 0 & -\ii \eps\, \nu^*(\rmvec{k}) \\
        0 & 0 & 0 \\
        \, \ii \eps\, \nu(\rmvec{k}) & 0 & 0 
    \end{bmatrix}
\end{equation}
where 
\begin{align}
  \ii \nu(\rmvec{k}) = &~ Q(\rmvec{w}_1,-\rmvec{w}_2)+ Q(-\rmvec{w}_1,\rmvec{w}_2)\ee^{\ii \rmvec{k}\cdot (\rmvec{v}_2-\rmvec{v}_1)} \\
    &+ Q(\rmvec{w}_1,\rmvec{w}_2)\ee^{\ii \rmvec{k}\cdot \rmvec{v}_2} 
    + Q(-\rmvec{w}_1,-\rmvec{w}_2)\ee^{-\ii \rmvec{k}\cdot\rmvec{v}_1 } \notag
\end{align}
and $Q$ is defined by equation (\ref{Eq: Qwv}). The nonzero terms in (\ref{Lieb_N_correction}) correspond to next-nearest neighbor interactions between the first and third lattice sites. 
The eigenvalue problem for approximating the spectrum of the Lieb lattice is now found to be 
\begin{equation}
    \lambda \bm\alpha_0+ 
        \begin{bmatrix}
            0 & \PAvg{\gamma_1}^*& -\ii \eps \nu^* \\
            \PAvg{\gamma_1} & 0 &      
                \PAvg{\gamma_2}^*\\
             \ii \eps \nu & \PAvg{\gamma_2}& 0 
        \end{bmatrix}
    \bm \alpha_0 = 0.
\end{equation}

\begin{figure}[ht]
    \centering
        \begin{subfigure}[b]{0.26\textwidth}
            \centering
            \includegraphics[width=\textwidth]{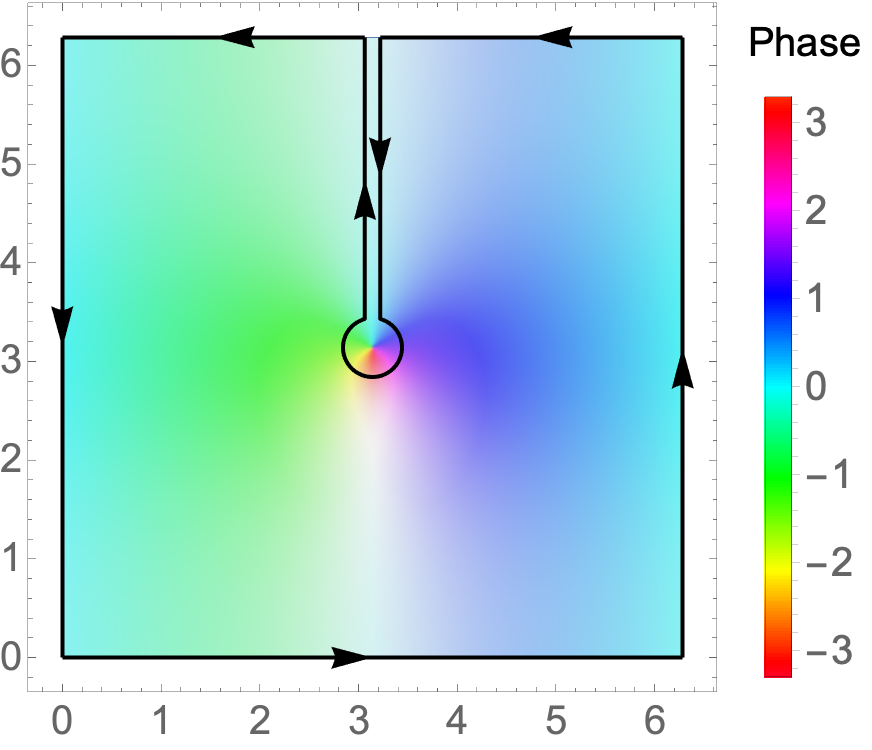}
            \caption{Magnitude and phase for the first component of $\bm \alpha_0$.} 
            \label{Fig: LiebAlpha1Top}
        \end{subfigure}
        \hspace{0.2cm}
        \begin{subfigure}[b]{0.265\textwidth}
            \centering
            \includegraphics[width=0.81\textwidth]{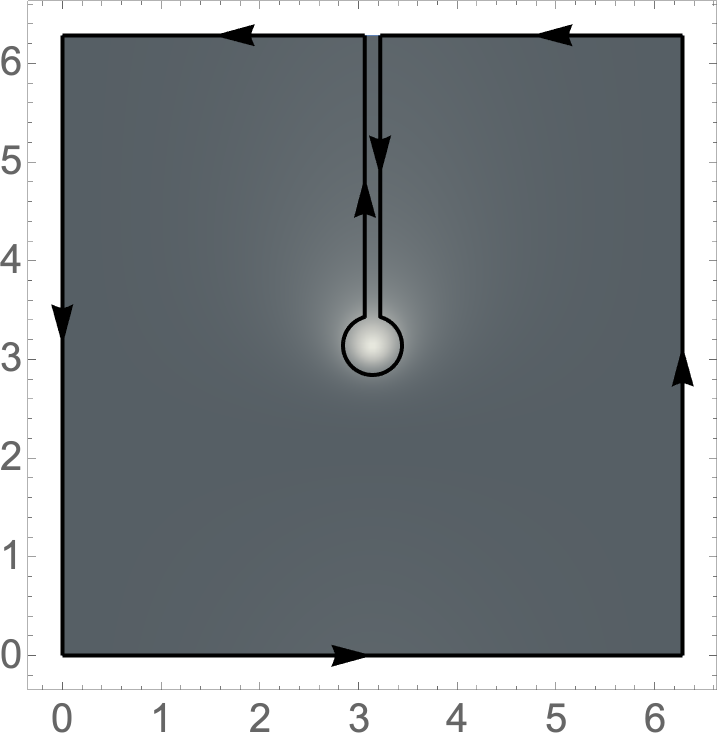}
            \caption{Magnitude and phase for the second component of $\bm \alpha_0$.}
            \label{Fig: LiebAlpha2Top}
        \end{subfigure}
        \hspace{0.2cm}
        \begin{subfigure}[b]{0.26\textwidth}
            \center
            \includegraphics[width=\textwidth]{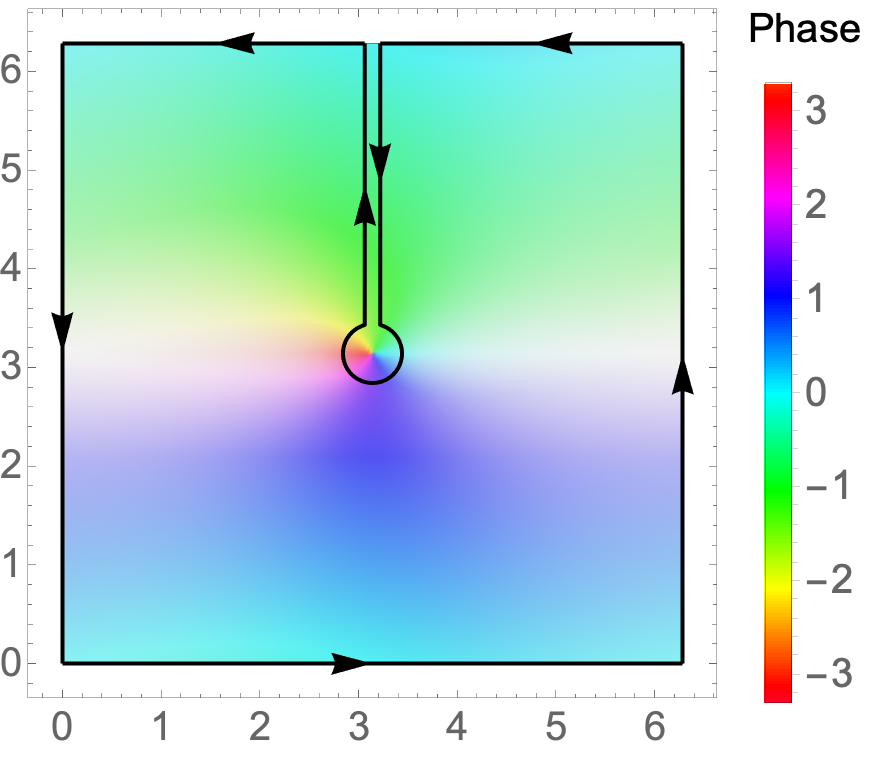}
            \caption{Magnitude and phase for third component of $\bm \alpha_0$.} 
            \label{Fig: LiebAlpha3Top}
        \end{subfigure}
    \caption{Contour of integration that removes the critical point at $\bm{K}_D = \left(\frac{\pi}{2},\frac{\pi}{2}\right)$. The magnitudes and phases are shown for the components of $\bm\alpha_0(\bf k)$ on the upper band. 
    The magnitudes are represented by color saturation (with white representing a magnitude of $0$) and the phases are represented by hue (with the second component of $\bm \alpha_0$ being strictly real and shown in grey scale).}
    \label{Fig: LiebContour}
\end{figure}

This systems has the eigenvalues 
\begin{equation}
   \label{lieb_spec}
   \lambda = 0, \pm \sqrt{\left|\PAvg{\gamma_1}\right|^2 + \left|\PAvg{\gamma_2}\right|^2 + \eps^2\left|\nu\right|^2 }.
\end{equation}
Note: The constant term in the characteristic polynomial, $\PAvg{\gamma_1}\PAvg{\gamma_2}\nu^*-\PAvg{\gamma_1}^*\PAvg{\gamma_2}^*\nu$, is 
zero and we find that $\PAvg{\gamma_1}\PAvg{\gamma_2}\nu^*$ is strictly real.
The eigenvectors can be expressed in terms of $\lambda(\rmvec{k})$ as
\begin{equation}
    \bm \alpha_0 = \disp\frac{1}{D(\rmvec{k})}
        \begin{bmatrix}
            \lambda\PAvg{\gamma_1}^* + \ii \eps \nu^* \PAvg{\gamma_2}\\
            -\lambda^2 +\eps^2 \left|\nu\right|^2\\
            \lambda \PAvg{\gamma_2} - \ii \eps \nu \PAvg{\gamma_1}^*
        \end{bmatrix}
        \label{Eq: LiebEigenvector}
\end{equation}
where $D(\rmvec{k})$ is a normalizing function 
\begin{align}
    D(\rmvec{k})^2 = &\lambda^2 (2\left|\PAvg{\gamma_1}\right|^2+2\left|\PAvg{\gamma_2}\right|^2-\eps^2\left|\nu\right|^2) \notag\\
    &+ \eps^2\left|\nu\right|^2(\left|\PAvg{\gamma_1}\right|^2+\left|\PAvg{\gamma_2}\right|^2+\eps^2\left|\nu\right|^2) 
\end{align}
which makes $\lVert \bm \alpha_0\rVert =1$.
In this case, we have chosen to fix the second component of $\bm\alpha_0$ as strictly real. 
When not on the flat band, i.e., $\lambda = 0$, $D$ further reduces to $D(\rmvec{k})^2=2\lambda^2(\left|\PAvg{\gamma_1}\right|^2+\left|\PAvg{\gamma_2}\right|^2)$. 
Notice that $D(\rmvec{k})$ vanishes when $\PAvg{\gamma_1}=\PAvg{\gamma_2}=0$, which corresponds to the Dirac point of the unperturbed spectrum, i.e., $\eps =0$. 
This has the effect of creating a discontinuity from the twist in the phase around the zeros of the first and third components of the normalized eigenvector (\ref{Eq: LiebEigenvector}) (see Figure \ref{Fig: LiebContour}). 
In terms of the magnitude, the zeros of $D$ are removable discontinuities and the important discontinuity appears only in the phase of the eigenvector. 
As such, there is a $2\pi$ change in the eigenfunction or so called Berry phase for a loop around the discontinuity. 

The discontinuity in the spectrum occurs at $\bm{K}_D = \left(\frac{\pi}{2},\frac{\pi}{2}\right)$, which means our calculation of the Chern number will be determined by the local behavior near this point. For $\rmvec{k} = \rmvec{K}_D + \delta\rmvec{ k}/q_0$ we have the approximations 
\begin{subequations}
\begin{align}
    \PAvg{\gamma_1}(\rmvec{k}) & \approx-2\ii\, \delta k_x\\
    \PAvg{\gamma_2}(\rmvec{k}) & \approx-2\ii\, \delta k_y\\
    \nu(\rmvec{k}) & \approx \nu(\rmvec{K}_D) \equiv \nu_0
\end{align}
\end{subequations}
where $\nu_0$ is a real nonzero constant. In this region the set of eigenvalues are approximated by  $\lambda(\rmvec{k} ) \approx \left\{ 0, \pm \sqrt{4 \delta k_x^2 + 4 \delta k_y^2 + \eps^2 \nu_0^2}\right\}$. In terms of their graphs, these spectral surfaces correspond to a flat plane and a hyperboloid of two sheets. On the upper and lower bands, the related eigenvectors are given by 
\begin{equation}
    \bm \alpha_0(\rmvec{k}) \approx \disp\frac{2}{D(\rmvec{k} )}
        \begin{bmatrix}
            -\ii  \delta k_x\lambda+ \delta k_y\eps\nu_0\\
            -2\left(\delta k_x^2 + \delta k_y^2\right)\\
            ~~\ii \delta k_y\lambda+\delta k_x\eps\nu_0
        \end{bmatrix}
\end{equation}
where $D(\rmvec{k})^2 \approx 8\left(\delta k_x^2 + \delta k_y^2\right)\lambda^2$.

From here, the approximation of the Berry connection is found to be
\begin{equation}
    \mathcal{A}= \mathcal{A}_1\hat{\rmvec{i}}+\mathcal{A}_2\hat{\rmvec{j}}
     \approx \disp\frac{\eps \nu_0\ii}{\lambda \left(\delta k_x^2 + \delta k_y^2\right) } \left( \delta k_y \hat{\rmvec{i}} - \delta k_x \hat{\rmvec{j}} \right)
    \label{Eq: LiebConnection}
\end{equation}
and the discontinuity in the eigenvectors is seen to produce a singularity in the Berry connection. 
Note: On the flat band where $\lambda =0$ the connection is zero. 
The approximation for the Berry curvature is then
\begin{equation}
    \Omega=(\partial_{k_x} \mathcal{A}_2-\partial_{k_y} \mathcal{A}_1)\uveck \approx \disp\frac{\ii 4 \eps \nu_0}{\lambda^3}\uveck.
\end{equation}
Unlike the eigenvectors and the Berry connection, the Berry curvature for the upper and lower bands are both smooth and bounded since $|\lambda| >0$. 
As a result, analogous to what was done in the honeycomb lattice calculation, we can remove an infinitesimal region around $\rmvec{k} = \rmvec{K}_D$ from the Brillouin zone without effecting the total curvature or Chern number. 
Once the singular point has been removed in this manner, Stokes' theorem can be employed to calculate the Chern number using the local behavior of the connection instead of the global behavior of the curvature. The radial symmetry of the differential $\bm{\mathcal{A}} \cdot \dd \rmvec{k}$ implies that removing a circular disk $\mathcal{D}(\delta) = \left\{\bm k\middle| \lVert \bm k - \bm K_D\rVert \leq \delta \right\}$ will facilitate the direct calculation of all relevant integrals.

The particular steps of this reasoning are outlined below
\begin{subequations}
\begin{align}
    C&= \disp\frac{1}{2\pi \ii} \disp\iint_{\mathcal{B}} \Omega(\rmvec{k}) \,\, \dd k_x \dd k_y \\
   % &= \disp\lim_{\delta \rightarrow 0} \disp\frac{1}{2\pi \ii} \disp\iint_{\mathcal{B}-\mathcal{D}(\delta)} \Omega(\rmvec{k}) \,\, \dd k_x \dd k_y \\
    &=  \disp\lim_{\delta \rightarrow 0} \disp\frac{1}{2\pi \ii} \disp\iint_{\mathcal{B}-\mathcal{D}(\delta)} \Big( \nabla_{\rmvec{k}} \times
    \bm{\mathcal{A}} \Big) \cdot \uveck\, \dd k_x \dd k_y \\
    & =  \disp\lim_{\delta \rightarrow 0} \disp\frac{1}{2\pi \ii}\left( \oint_{\partial\mathcal{B}}\mathcal{A}(\rmvec{k}) \cdot % \,
     \dd \bm k - \disp\oint_{\partial \mathcal{D}(\delta)} \mathcal{A}(\rmvec{k})  \cdot  % \,
      \dd \bm k\right)\\
    & = \disp\frac{1}{2\pi \ii} \disp\lim_{\delta \rightarrow 0}  \left(-\disp\oint_{\partial \mathcal{D}(\delta)} \mathcal{A}(\bm k) \cdot %\, 
    \dd \bm k\right)\\
    & = \disp\frac{1}{2\pi }\disp\lim_{\delta \rightarrow 0}\left(-\sign(\lambda)\int_0^{2\pi} \disp\frac{\eps\nu_0}{\sqrt{4\delta^2 + (\eps \nu_0)^2}}\dd\,\theta
    \right)\\
    & = \disp\lim_{\delta \rightarrow 0}\left(-\sign(\lambda) \disp\frac{\eps\nu_0}{\sqrt{4\delta^2 + (\eps \nu_0)^2}}\right)\\
    & = -\sign(\lambda \nu_0),
\end{align}
\end{subequations}
where we have used the periodicity of the Berry connection along the boundary of the Brillouin zone. 
The end result is a pair of opposite Chern numbers of $\pm 1$ for the top and bottom bands. Numerical calculation of coefficients in the function $\nu$ shows that $\nu_0 >0$ for a wide range of parameters and the upper band has a Chern number $-1$ while the bottom band has a Chern number of $+1$. The flat middle band has a Chern number of $0$, which can be found by direct application of Stokes' theorem since there are no discontinuities present in the Berry connection. This is in agreement with numerical calculations of the spectrum in \cite{Cole2019}.

%%%%%%%%%%%%%%%%%%%%%%%%%%%%%%%%% KAGOME %%%%%%%%%%%%%%%%%%%%%%%%%%%%%%%%%%%%%%%%

\subsection{Kagome Lattice}
\label{Sec: Kagome}

\begin{figure}[ht]
    \centering
        \begin{subfigure}[b]{0.65\textwidth}
            \centering
            \includegraphics[width=0.65\textwidth]{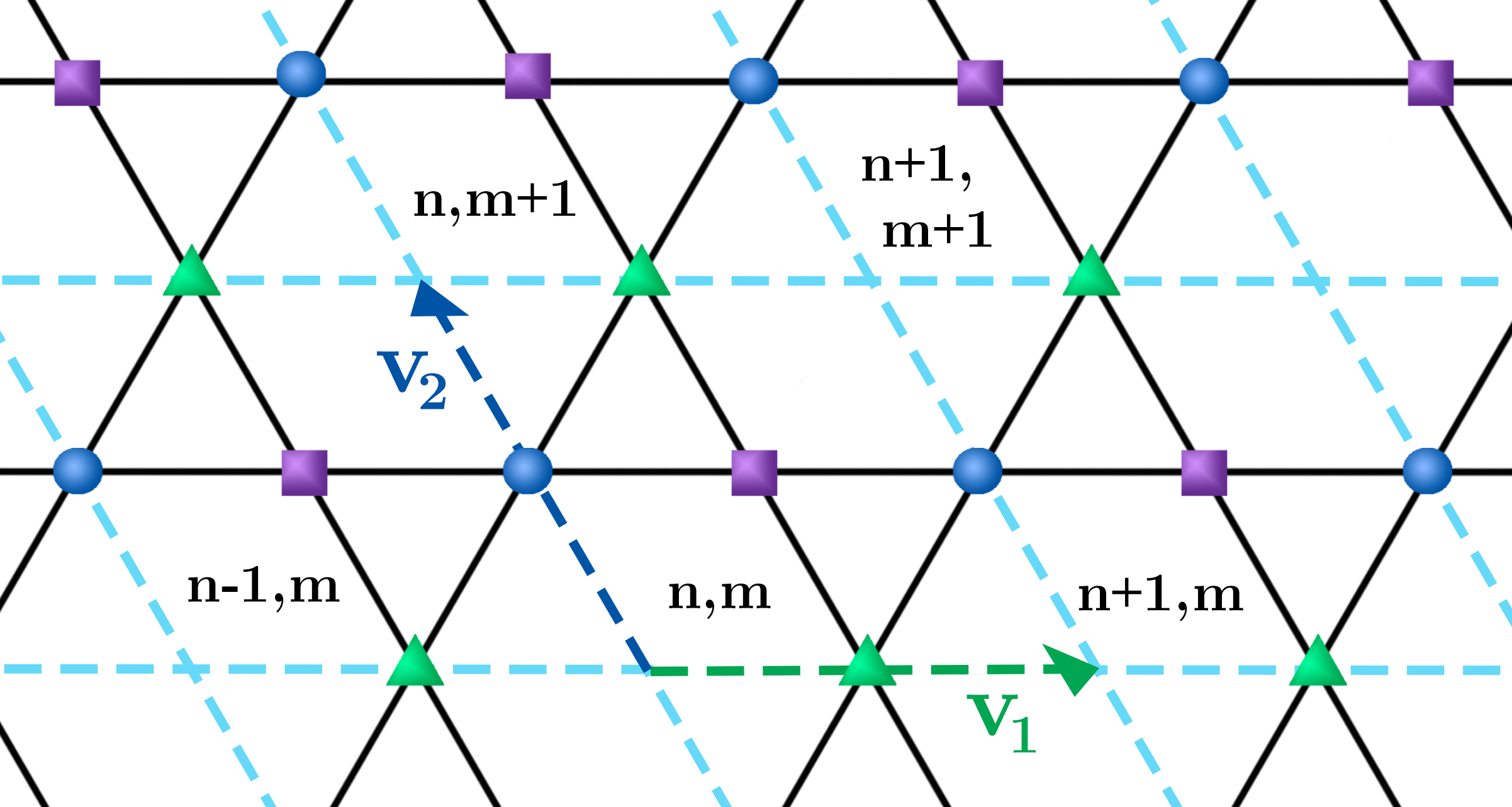}
             \hspace{0.1cm}
            \includegraphics[width=0.3\textwidth]{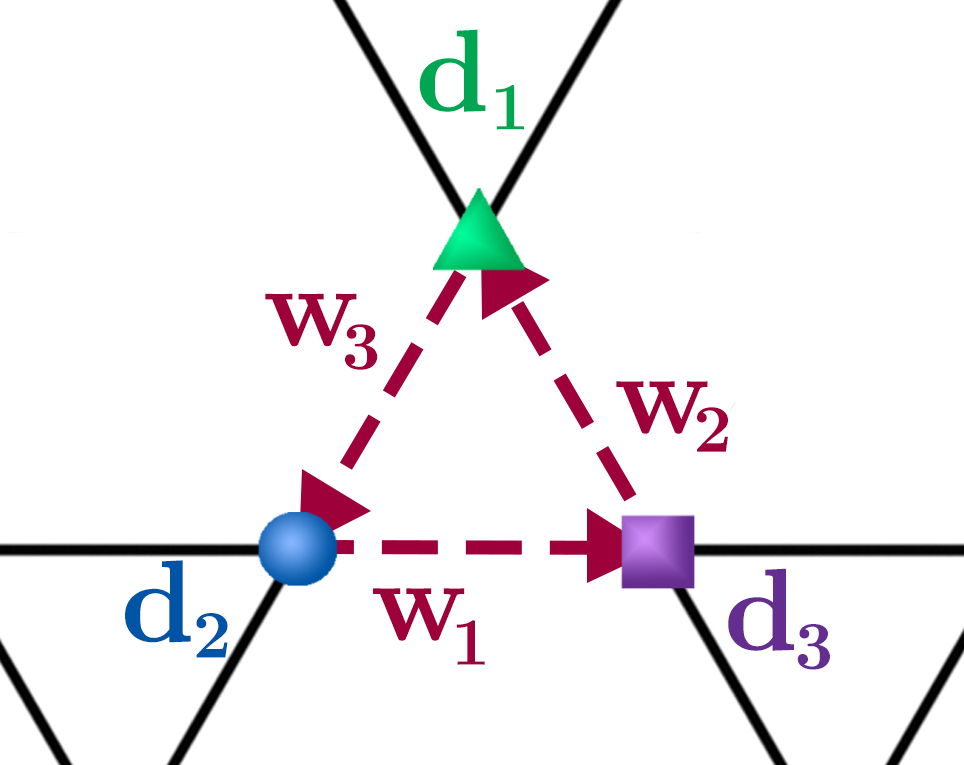}
                        \caption{Diagram for the Kagome lattice and except for a single lattice cell.}
        \end{subfigure}
        \begin{subfigure}[b]{0.29\textwidth}
        		\center
                \includegraphics[width=0.9\textwidth]{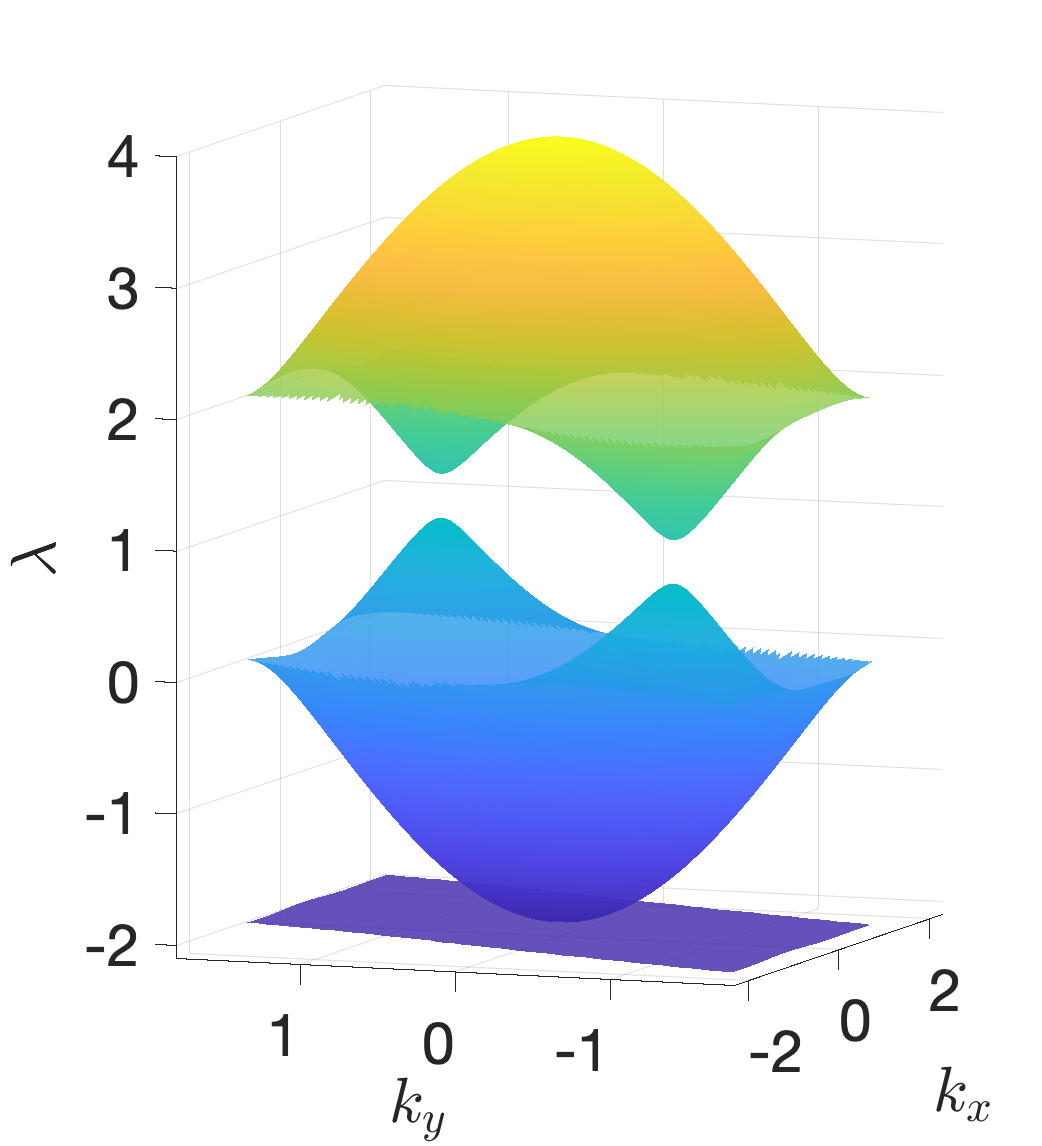}
                \caption{Spectrum for the kagome lattice.}
                \label{Fig: KagomeSpectrum}
        \end{subfigure}
    \caption{The kagome kattice consists of three sublattices. The green triangles represent the first sublattice, the blue spheres represent the second sublattice and the purple squares represent the third sublattice.}
    \label{Fig: KagomeLattice}
\end{figure}

Here we consider the kagome lattice (see Figure \ref{Fig: KagomeLattice}) as an example \cite{Kagome2009}. In a similar way to how the Lieb lattice is a 1/4-depleted square lattice, the kagome lattice is a triangular lattice where one in every four lattice sites has been removed. 
In Figure \ref{Fig: KagomeLattice}, the lattice cell is drawn so that the corners are located where the lattice sites would have been removed from a triangular lattice. When considering the behavior in the Brillouin zone the, the kagome system features two Dirac points between the middle and top bands as well as a third parabolic degeneracy between the nearly flat bottom band and the middle band (see Figure \ref{Fig: KagomeSpectrum}). While the discontinuities of the eigenfunctions bifurcate from the Dirac points for small oscillations ($\kappa \ll 1 $), they do not remain stationary as is the case for honeycomb and Lieb lattices. 

In terms of the framework laid out in Section \ref{Sec: Model}, the kagome lattice is defined by $\vv_1 = (2,0)$, $\vv_2 = (-1,\sqrt{3})$  (though not used to define the lattice cell we define $\vv_3 = (-1,-\sqrt{3})$) and $\rmvec{d}_1 = \left(0,\sqrt{3}\right)$, $\rmvec{d}_2 = \left(-\frac{1}{2},\frac{\sqrt{3}}{2}\right)$, $\rmvec{d}_3 = (\frac{1}{2},\frac{\sqrt{3}}{2})$. From here we can derive the nearest neighbor vectors $\rmvec{w}_1 = (1,0)$, $\rmvec{w}_2 = \left(-\frac{1}{2},\frac{\sqrt{3}}{2}\right)$, and $\rmvec{w}_3 = \left(-\frac{1}{2},-\frac{\sqrt{3}}{2}\right)$. The system of differential equations, in the form (\ref{Eq: MatrixODE}), derived for this lattice is
\begin{equation}
    \ii \frac{\dd \bm{\alpha}}{\dd z}+ 
        \begin{bmatrix}
            0 & \gamma_3^*(z,\rmvec{k})& \gamma_2(z,\rmvec{k}) \\
            \gamma_3(z,\rmvec{k}) & 0 & \gamma_1^*(z,\rmvec{k})\\
            \gamma_2^*(z,\rmvec{k}) & \gamma_1(z,\rmvec{k})& 0 
        \end{bmatrix}
    \bm \alpha = 0
    \label{Eq: KagomeFull}
\end{equation}
where the $\gamma_j$ and their averages $\PAvg{\gamma_j}$ functions are given by
\begin{subequations}
\begin{align}
\gamma_j(z , \rmvec{k})  &=  q(z,\rmvec{w}_j) \ee^{\ii \rmvec{k}\cdot\rmvec{v}_j}+q^*(z,\rmvec{w}_j) \\
\PAvg{\gamma_j}(\rmvec{k})  &=  q_0 \Big(\ee^{\ii \rmvec{k}\cdot\rmvec{v}_j}+1 \Big) 
\end{align}
\end{subequations}
for the function $q$ previously defined in (\ref{Eq: qz}). Looking at Figure \ref{Fig: KagomeLattice}, we see that all interactions come in pairs pointing in opposite directions. 

Plugging (\ref{Eq: KagomeFull}) into the general averaged equation (\ref{Eq: AveragedSystem}) yields
\begin{equation}
    \PAvg{M}(\rmvec{k})= \begin{bmatrix}
            0 & \PAvg{\gamma_3}^*& \PAvg{\gamma_2}~ \\
            \PAvg{\gamma_3}~ & 0 & \PAvg{\gamma_1}^*\\
            \PAvg{\gamma_2}^* & \PAvg{\gamma_1}~& 0 
        \end{bmatrix}.
\end{equation}
and at order $\eps$, the correction term is
\begin{equation}
\PAvg{\Nu}(\rmvec{k})  = 
   \begin{bmatrix}
        0 & \nu_{21}(\rmvec{k}) &  \nu_{31}(\rmvec{k}) \\
         \nu_{12}(\rmvec{k}) & 0 &  \nu_{32}(\rmvec{k})  \\
        \nu_{13}(\rmvec{k}) &  \nu_{31}(\rmvec{k}) & 0 
    \end{bmatrix}
\end{equation}
where 
\begin{equation*}
    \nu_{ij}(\rmvec{k}) = \ee^{-\ii \rmvec{k} \cdot (\rmvec{d}_j + \rmvec{d}_i)}\Bigg( Q(\rmvec{w}_i,\rmvec{w}_j) \ee^{\ii \rmvec{k} \cdot (\rmvec{w}_i + \rmvec{w}_j)} + Q(-\rmvec{w}_i,\rmvec{w}_j) \ee^{\ii \rmvec{k} \cdot (-\rmvec{w}_i + \rmvec{w}_j)}-c.c.\Bigg)
    % \\
    % &+Q(\rmvec{w}_i,-\rmvec{w}_j) \ee^{\ii \rmvec{k} \cdot (\rmvec{w}_i - \rmvec{w}_j)} + Q(-\rmvec{w}_i,-\rmvec{w}_j) \ee^{\ii \rmvec{k} \cdot (-\rmvec{w}_i -\rmvec{w}_j)}\Bigg)
    % \\
    % \equiv& -\ii \ee^{-\ii \rmvec{k} \cdot (\rmvec{d}_j + \rmvec{d}_i)} \Big(\ii Q(\rmvec{w}_i,\rmvec{w}_j) \ee^{\ii \rmvec{k} \cdot (\rmvec{w}_i + \rmvec{w}_j)} + \ii Q(-\rmvec{w}_i,\rmvec{w}_j) \ee^{\ii \rmvec{k} \cdot (-\rmvec{w}_i + \rmvec{w}_j)}+ C.C.\Big)
\end{equation*}
and ``c.c." stands for complex conjugate. We can formulate the averaged eigenfunctions and the corresponding connection, curvature and the Chern numbers using the same framework from Section \ref{Sec: Lieb}.
The Chern numbers are found numerically (using the algorithm in \cite{Fukui2005}) to be $-1$, $0$, and $1$ for the bottom, middle, and top band respectively
; this agrees with the results in \cite{Cole2019}.\\

%%%%%%%%%%%%%%%%%%%%%%%%%%%%%%%%% One Fifth %%%%%%%%%%%%%%%%%%%%%%%%%%%%%%%%%%%%%%%%

\subsection{1/5 Depleted Lattice}
\label{Sec: OneFifth}

\begin{figure}[ht]
    \centering
        \begin{subfigure}[b]{0.52\textwidth}
            \centering
            \includegraphics[width=0.6\textwidth]{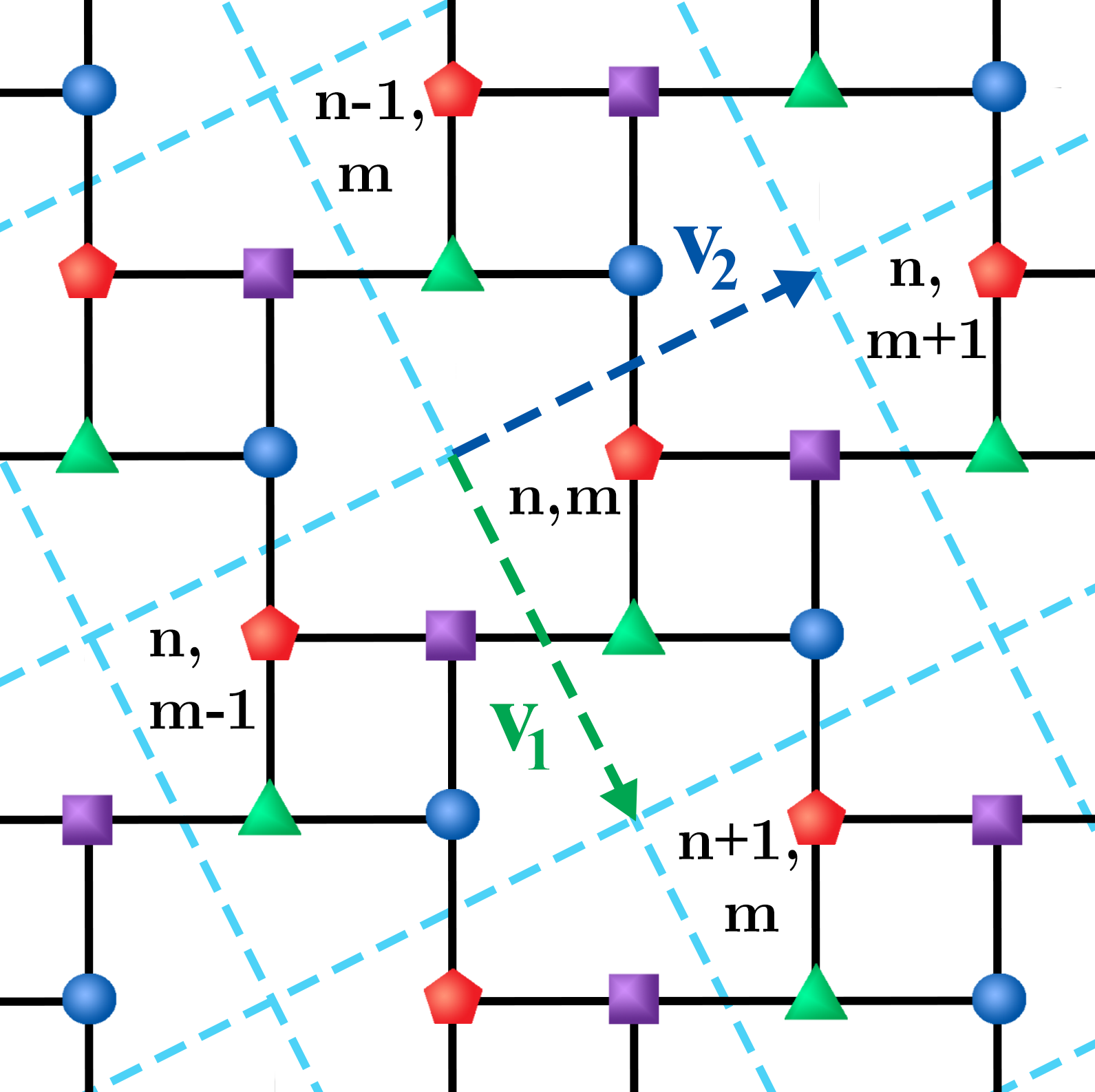} \hspace{0.1cm}
            \includegraphics[width=0.33\textwidth]{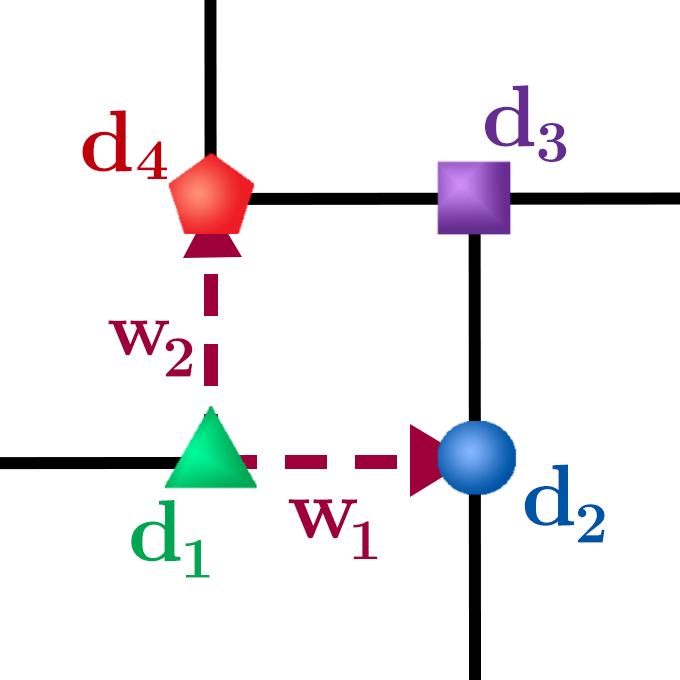}
            \caption{Diagram for the 1/5 depleted lattice with four lattice sites per unit cell.}
            %\vspace{0.3cm}
        \end{subfigure}
        \hspace{0.2cm}
        \begin{subfigure}[b]{0.32\textwidth}
    \centering
        \includegraphics[width=0.9\textwidth]{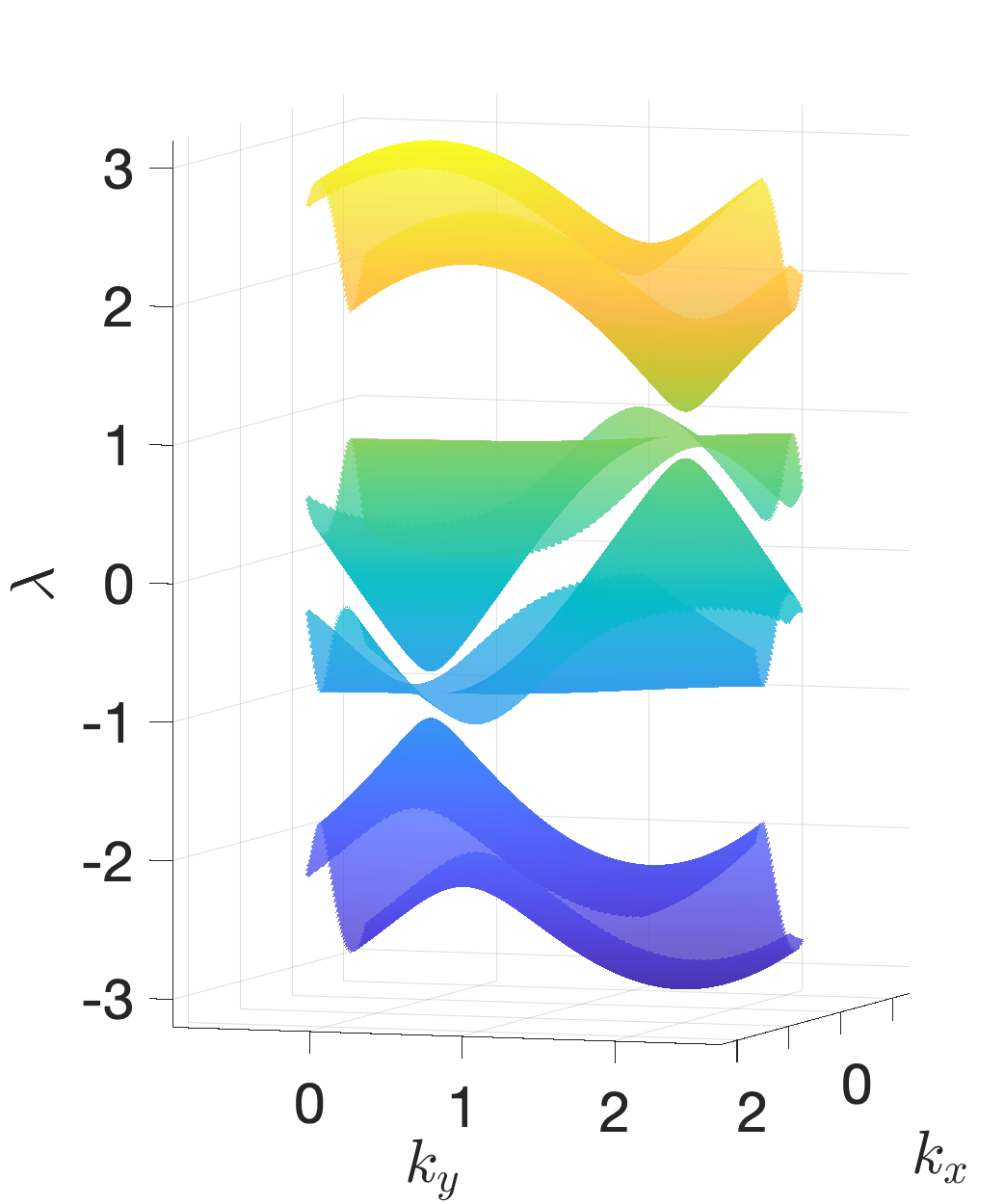}
    \caption{Spectrum for the 1/5 depleted lattice.}
            \vspace{0.3cm}
        \end{subfigure}
    \caption{The 1/5-depleted Lattice consists of four sublattices. The green triangles represent the first sublattice, the blue squares 
    represent the second sublattice, the purple squares represent the third sublattice, and the red pentagrams represent the fourth sublattice. }
    \label{Fig: OneFifthLattice}
\end{figure}

Here we consider the 1/5 depleted lattice \cite{Manuel1998} as an example. As a 1/5 depleted lattice, one in every five lattice sites is being removed (see e.g. Figure \ref{Fig: OneFifthLattice}) from a square lattice. The choice of lattice cell is chosen such that the corners of the cell lie at the location of the removed lattice sites. The spectrum for the 1/5 depleted lattice %(see \ref{Fig: OneFifthLattice}) 
has two Dirac points similar %akin
to the kind found for the Lieb lattice in where a flat band sits between upper and lower sections of the conical spectrum. In this case, the middle band is only locally flat near the Dirac point, but not %but not 
across the entire Brouillon zone. Though they do not touch, there is no gap in $\lambda$ between the lower middle and upper middle bands. 

In terms of the framework laid out in Section \ref{Sec: Model}, the 1/5 depleted lattice is defined by $\vv_1 = (1,-2)$, $\vv_2 = (2,1)$ and $\rmvec{d}_1 = (1, -1)$, $\rmvec{d}_2 = (2,-1)$, $\rmvec{d}_3 = (2,0)$, $\rmvec{d}_4  = (1,0)$. From here we can derive the nearest neighbor vectors $\rmvec{w}_1 = (1,0)$ and $\rmvec{w}_2 = (0,1)$. The system of differential equations, in the form (\ref{Eq: MatrixODE}), derived for this lattice is 
\begin{equation}
    \ii \frac{\dd \bm{\alpha}}{\dd z}+ 
        \begin{bmatrix}
         0 & q(z,\rmvec{w}_1)& \gamma_1^*(z,\rmvec{k})& q(z,\rmvec{w}_2)\\
         q^*(z,\rmvec{w}_1)& 0 & q(z,\rmvec{w}_2)&\gamma_2^*(z,\rmvec{k})\\
         \gamma_1(z,\rmvec{k})& q^*(z,\rmvec{w}_2)& 0&q^*(z,\rmvec{w}_1)\\
         q^*(z,\rmvec{w}_2)& \gamma_2(z,\rmvec{k})& q(z,\rmvec{w}_1)&0
        \end{bmatrix}
    \bm \alpha = 0
    \label{Eq: OneFifthFull}
\end{equation}
where the $\gamma_j$ functions are given explicitly by
\begin{subequations}
\begin{align}
    \gamma_1(z , \rmvec{k}) & = q(z,\rmvec{w}_1) \ee^{\ii \rmvec{k}\cdot\rmvec{v}_2}\\
    \gamma_2(z , \rmvec{k}) & = q(z,\rmvec{w}_2) \ee^{-\ii \rmvec{k}\cdot\rmvec{v}_1}
\end{align}
\end{subequations}
and we note that $q^*(z,\rmvec{w}) = q(z,-\rmvec{w})$. 

It is often a feature of more complicated lattices that there are few interactions between sublattices.
Here for example, each of the four sublattices of the 1/5 depleted lattice has only one nearest neighbor on each of the other sublattices, but there were two interactions in the cases of Lieb and kagome lattices and three for the hexagonal lattice. This means the individual elements of $M$ become simpler as the order of $M$ increases.

Substituting (\ref{Eq: OneFifthFull}) into the general averaged equation (\ref{Eq: AveragedSystem}) gives 
\begin{equation}
    \PAvg{M}(\rmvec{k})= q_0
        \begin{bmatrix}
            0 & 1 & \ee^{-\ii\rmvec{k}\cdot\rmvec{v}_2} & 1  \\
            1 & 0 & 1 & \ee^{\ii\rmvec{k}\cdot\rmvec{v}_1}  \\
            \ee^{ \ii \rmvec{k}\cdot\rmvec{v}_2} & 1 & 0 & 1 \\
            1 & \ee^{-\ii \rmvec{k}\cdot\rmvec{v}_1} & 1 & 0  \\
        \end{bmatrix}.
\end{equation}
At order $\eps$, the correction term is
\begin{equation}
\PAvg{\Nu}(\rmvec{k})  = 
    \begin{bmatrix}
        \nu_0 & 
    -\nu_1^*~\ee^{-\ii \rmvec{k} \cdot \rmvec{v}_2} &
    0 &
    \nu_1^*~\ee^{\ii \rmvec{k} \cdot \rmvec{v}_1} \\
        
    -\nu_1 \ee^{\ii \rmvec{k} \cdot \rmvec{w}_2} &
    0 & -\nu_1^* \ee^{\ii \rmvec{k} \cdot \rmvec{v}_1}&0 \\
     
    0& - \nu_1\ee^{-\ii \rmvec{k} \cdot \rmvec{v}_1}
    &- \nu_0 & 
    \nu_1 \ee^{\ii \rmvec{k} \cdot \rmvec{v}_2} \\
     
    \nu_1 \ee^{-\ii \rmvec{k} \cdot \rmvec{v}_1}  & 0 & \nu_1^*\ee^{-\ii \rmvec{k} \rmvec{v}_2} &
    0 
    \end{bmatrix}
\end{equation}
where $\nu_0 = 2 q_0 s_0 \kappa$ and $\nu_1 = \frac{1}{2}\big(\nu_0 + \ii s_0^2 \kappa^2\big)$. As before, using the averaging framework and the algorithm in \cite{Fukui2005} the Chern numbers are found as $1$, $1$, $-1$ and $-1$ for the bands running from %for 
bottom to top. Due to space limitations we omit intermediate details.

\section{Nonlinear Envelope Equations}

From the linear analysis, we have found that even though the Chern Number is a global property of the spectrum, it can be successfully calculated from the local behavior near critical points in the spectrum under suitable restrictions on the eigenvectors. We now consider the nonlinear dynamics of wave packets in the gap near the critical points of the linear spectrum. For this, we assume weak nonlinearity.
\begin{equation}
    \ii \disp \frac{\partial \bm\alpha}{\partial z } + M\left( z,\rmvec{k} \right)\bm\alpha + \varepsilon \widetilde{\sigma} \mathbb{N}(\bm \alpha)\bm \alpha= 0.
    \label{Eq: MatrixODENL}
\end{equation}
where the nonlinear term once again is defined as $\mathbb{N}(\bm \alpha) = \mathrm{diag}(|\alpha^1|^2,\ldots,|\alpha^{\ell}|^2)$.

Following the steps from Section \ref{Sec: Averaged} the averaged nonlinear system when the  nonlinear term first appears is found to be
\begin{equation}
    \ii \pdv{\bm{\alpha}}{z} +  \left(\left<M\right>(\rmvec{k}) + \eps \PAvg{\Nu}(\rmvec{k})\right) \bm\alpha + \eps \widetilde{\sigma} \mathbb{N}(\bm{\alpha}) \bm{\alpha}= 0.
    \label{Eq: AveragedSystemNL}
\end{equation}
Note that we are assuming a balance between the nonlinearity and the higher order term due to the averaging.
% ($\PAvg{\Nu}(\rmvec{k})\right)$.}

Near a critical point of the linear spectrum, $\rmvec{k} =\rmvec{K}_D +\eps \rmvec{k} $, we can expand the linear terms as 
\begin{equation}
    \PAvg{M} + \eps \PAvg{\Nu}\approx \PAvg{M}( \rmvec{K}_D) + \eps \rmvec{k} \cdot \nabla_{\rmvec{k}} \PAvg{M} + \eps \PAvg{\Nu}(\rmvec{K}_D) + \ldots
\end{equation}
where the gradient of the matrix $\PAvg{M}$ is applied element-wise.

To find an equation for the slowly-varying envelope of a wave packet, we expand the solution in form
\begin{equation}
\bm \alpha = \left(\bm\alpha_0\left(Z,\eps\rmvec{k}\right) + \eps \bm\alpha_1\left(Z,\eps\rmvec{k}\right) + \eps^2 \bm\alpha_2\left(Z,\eps\rmvec{k}\right) + \ldots\right) \ee^{-\ii \lambda_0 z}.
\end{equation}
where $Z=\eps z$ is a slow propagation variable. At leading order, $\bm \alpha_0$ satisfies the eigenvalue problem
\begin{equation}
    \lambda_0\bm\alpha_0 + \PAvg{M}(\rmvec{K}_D)\bm\alpha_0 = 0.
    \label{Eq: AveragedUnpert}
\end{equation}
Equation (\ref{Eq: AveragedUnpert}) represents the unperturbed, $\eps =0$, lattice at the critical point, $\rmvec{K}_D$, corresponds to a Dirac point of the linear spectrum. 
%Thus, there are at least two 
We assume linearly independent solutions (two in the case of Dirac , three in the case of the Lieb and kagome lattices). 
%This suggests the 
Superposition of $L$ slowly-varying envelopes, one for each eigenvector, yields
\begin{equation}
    \bm \alpha_0\left(Z, \eps\rmvec{k}\right)  = \sum_{i = 1}^{L} \widehat{E}_i\left(Z, \eps\rmvec{k}\right) \bm \alpha^{i}_0\left(\eps\rmvec{k}\right).
\end{equation}

At $\mathrm{O}(\eps)$, the nonlinear term balances against the linear terms of the band approximation
\begin{align}
    \lambda_0 \bm \alpha_1 + \PAvg{M}&(K_D)\bm\alpha_1 =\notag \\ -\sum_{i=1}^{L}\Bigg( &\ii \pdv{\widehat{E}_i}{Z}\bm \alpha_0^{i} + \rmvec{k} \cdot \nabla \PAvg{M}(\rmvec{ K}_D)\widehat{E}_i\bm \alpha_0^{i} + \PAvg{\Nu}(\rmvec{K}_D)\widehat{E}_i \bm \alpha_0^{i}
    \Bigg)\\& - \widetilde{\sigma}\mathbb{N}(\bm \alpha_0)\bm \alpha_0.\notag
\end{align}
This represents a maximal balance between not only the slowly varying envelope, the slow local dispersion  of the unperturbed (non-driven) lattice, and the strength of the nonlinearity, but also the warping of the dispersion relation due to the driving of the lattice. %This means 
Hence, we are either finding solutions that exist in the gap, as is the case for the honeycomb and kagome lattices, or  on the flat band, as is the case for the Lieb and 1/5 depleted lattices. As mentioned previously, $\PAvg{M}$ is a Hermitian matrix and so the Fredholm alternative says the right hand side must be orthogonal to the $L$ eigenvectors. 

%For the two examples from this paper,  
In the case of the honeycomb and Lieb lattices, things are further reduced since the eigenvectors span the whole space at the point of degeneracy. We assume  $\bm \alpha_0$ 
 can be written as a vector of envelopes, $\left(\widehat{E}_1(Z),\ldots, \widehat{E}_L(Z)\right)^T$, and after applying orthogonality the right hand side yields the following system of  nonlinear system of equations
 %must be equal to zero.. This agrees with the eigenvectors defined in section \ref{Sec: Chern}. In this case, we arrive the system of $L$ equations 
 for $\widehat{\rmvec{E}}= \left(\widehat{E}_1, \ldots, \widehat{E}_{L}\right)^T$
\begin{equation}
    \ii \pdv{\widehat{\rmvec{E}}}{Z} + \mathbb{L}\left(\rmvec{k}\right) \widehat{\rmvec{E}} + \widetilde{\sigma}\mathbb{N}\left(\widehat{\rmvec{E}}\right)\widehat{\rmvec{E}}=0
\end{equation}
where
\begin{equation}
    \mathbb{L}(\rmvec{k}) = \nabla_{ \rmvec{k}} \PAvg{M}(\rmvec{K}_D)\cdot {\bm k} + \PAvg{\Nu}(\rmvec{K}_D).
\end{equation}
%\begin{equation}
 %   \mathbb{L}(\bm k) = \nabla_{\bm k} \PAvg{M}(\rmvec{K}_D)\cdot \rmvec{k} + \PAvg{\Nu}(\rmvec{K}_D).
% \end{equation}
%%
Finally, moving back to physical space, in the neighborhood of $K_D$ we replace $\rmvec{k}$ with $-\ii \nabla_\rmvec{R}$ for the slow spatial variable $\rmvec{R} = \eps \rmvec{r}$.
\begin{equation}
    \ii \pdv{\rmvec{E}}{Z} + \mathbb{L}(-\ii \nabla_{\rmvec{R}}) \rmvec{E}+ \widetilde{\sigma}\mathbb{N}(\rmvec{E})\rmvec{E}=0.
\end{equation}
It should be noted that we have treated the nonlinear terms independently of the liner terms when transforming between the physical and wave number domains. 
This is a natural approach when the dispersive and nonlinear terms do not interact and results in the same final envelope equations as a more rigorous accounting of the analysis; see \cite{AblowitzRedBook} for more details on this type of argument. 

For the honeycomb lattice, the envelope equations are
\begin{subequations}
\begin{align}
    \ii \partial_Z E_1 -\frac{3}{2}(-\partial_X E_2 &+ \ii \partial_Y E_2)+ \nu_0 E_1+\widetilde{\sigma} \left|E_1\right|^2 E_1=0\\ 
    \ii \partial_Z E_2 -\frac{3}{2}(\partial_X E_1 &+\ii \partial_Y E_1)- \nu_0 E_2+\widetilde{\sigma} \left|E_2\right|^2 E_2=0
\end{align}
    \label{Eq: NLDirac}
\end{subequations}
where $\nu_0= \nu(K_D)$ and we assume a perfect honeycomb lattice $\rho=1$. %here $\rho=1$. 
This is the %familiar 
nonlinear Dirac system studied earlier cf. \cite{Nixon2009} but now with an important additional linear term proportional to $\nu_0$ that opens a gap in the dispersion relation sometimes called the Dirac equation with varying mass \cite{ ZhuDirac2022}.  The corresponding dispersion relations for plane waves $\exp\left[ i (K X + L Y - \Omega Z) \right]$ are
\begin{equation}
\Omega(K,L) = \pm \sqrt{ \nu_0^2 + \frac{9}{4} \left( K^2 + L^2 \right) }.
\end{equation}
Note: $\Omega(K,L)$ has a dispersion relation similar to that of two-dimensional Klein-Gordon equations. As such, there is dispersion and hence localized initial data will disperse for large $z$ \cite{AblowitzRedBook}. 
This spectrum is the continuous analog of the discrete eigenvalues in equation (\ref{Fig: HCSpectrum}), with a spectral gap at $(K,L) = (0,0)$ of width $2 |\nu_0|$.

\begin{figure}[ht]
    \centering
            \includegraphics[width=0.8\textwidth]{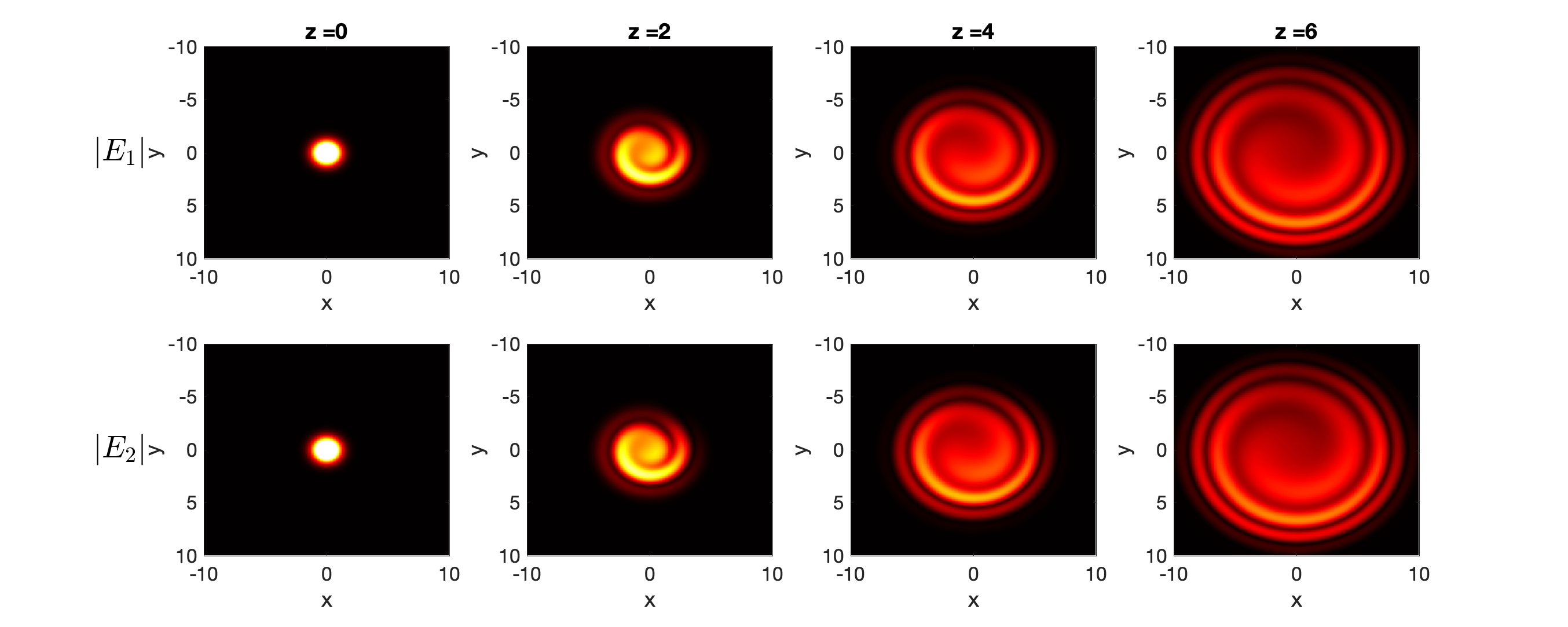}
            \caption{Evolution of an initial Gaussian pulse in the nonlinear Dirac system (\ref{Eq: NLDirac}) for $\nu_0 = 2$ and $\widetilde{\sigma} = 0.1$. The inclusion of the term $\nu_0$ term creates a spiral diffraction pattern.} 
            \label{Fig: NLDiracWeak}
\end{figure}

\begin{figure}[ht]
    \centering
            \includegraphics[width=0.8\textwidth]{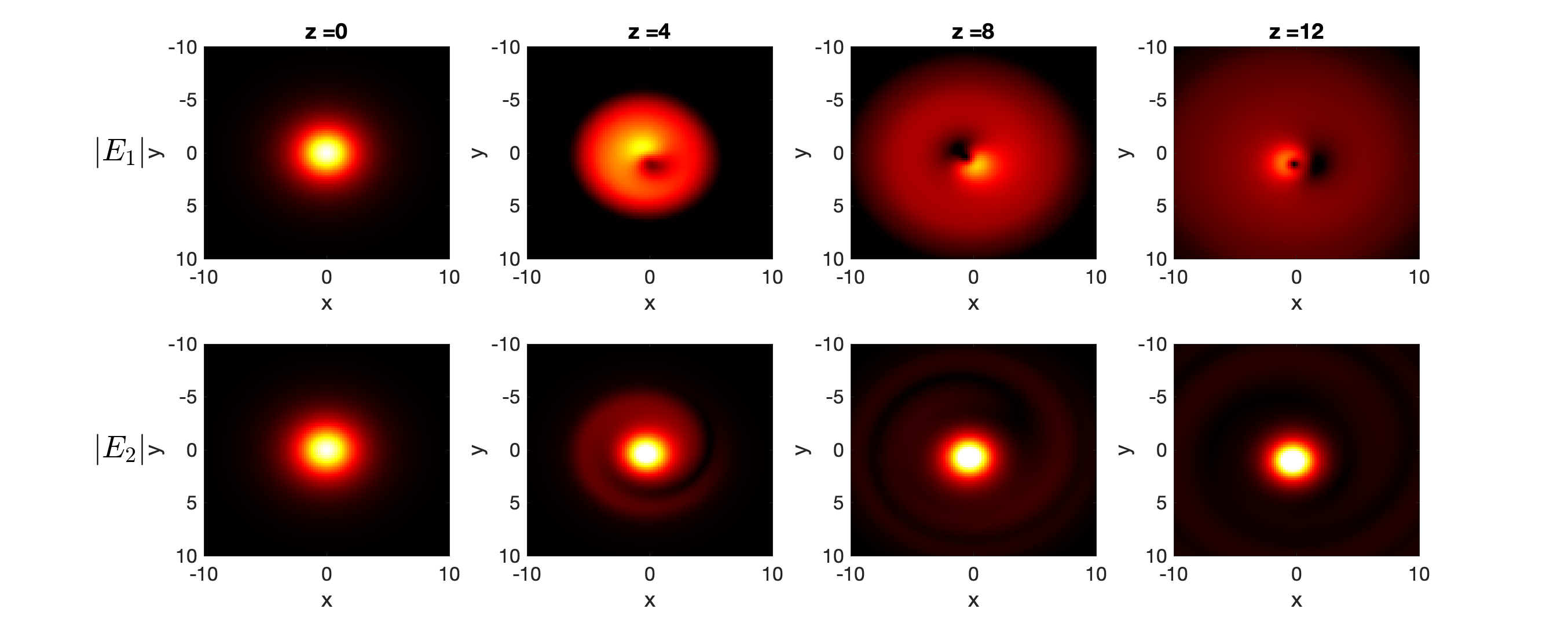}
            \caption{Evolution of the nonlinear Dirac system (\ref{Eq: NLDirac}) for $\nu_0 = 2$ and $\widetilde{\sigma} = 0.75$. A persistent, soliton-like, pulse develops in the second component.}
            \label{Fig: NLDiracStrong}
\end{figure}

\begin{figure}[ht]
    \centering
            \includegraphics[width=0.8\textwidth]{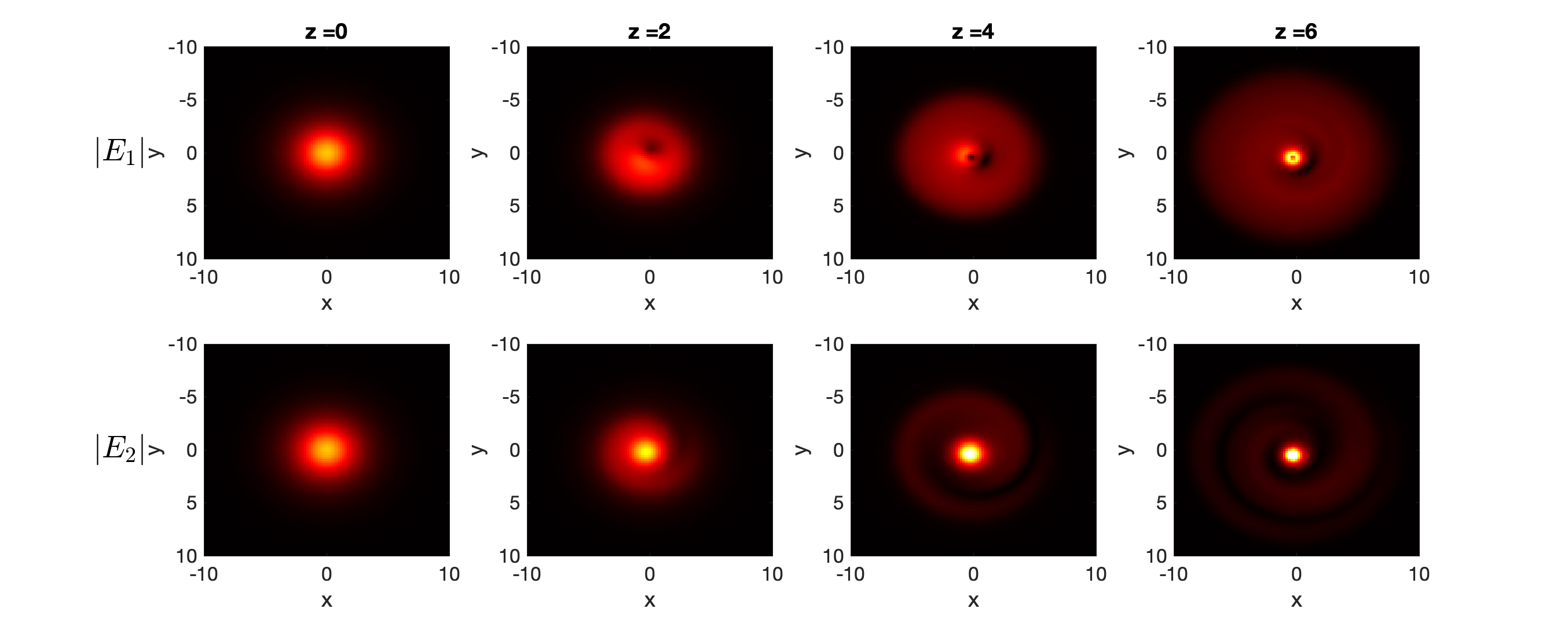}
            \caption{Evolution of an initial Gaussian pulse in the nonlinear Dirac system (\ref{Eq: NLDirac}) for $\nu_0 = 2$ and $\widetilde{\sigma} = 1$. For nonlinear above the critical threshold the pulse will undergo collapse in finite time.}
            \label{Fig: NLDiracCollapse}
\end{figure}

For the system (\ref{Eq: NLDirac}) we consider Gaussian initial envelopes over both carrier eigenfunctions.
The effect of the lattice rotation is to turn the conical diffraction usually associated with Dirac points \cite{Zhu2010}, into the spiral diffraction appearing in Figure \ref{Fig: NLDiracWeak}. Here the nonlinear effects are relatively weak, $\widetilde{\sigma}=0.10$, in comparison to the effects of lattice rotation, $\nu_0=2$. As the strength of the nonlinearity is increased, the nonlinear Dirac system transitions to a regime with wave collapse. This suggests the existence of soliton-like solution which persist over a long distance at some critical strength of the nonlinearity.
Figure \ref{Fig: NLDiracStrong} shows an example of such a soliton-like solution and we see that the light focus in the second second component of the system.
Figure \ref{Fig: NLDiracCollapse} shows that as nonlinearity further increases both components collapse. \\

%\mja{DOES CHANGING $\sigma$  FROM POSITIVE TO NEGATIVE MATTER RE: COLLAPSE?}

For the Lieb lattice, the envelope equations are
\begin{subequations}
\begin{align}
    \ii \partial_Z E_1 &- \partial_X E_2 -\ii \nu_0E_3+ \widetilde{\sigma} \left|E_1\right|^2E_1 =0\\
    \ii \partial_Z E_2 &+ \partial_X E_1 - \partial_Y E_3 + \widetilde{\sigma} \left|E_2\right|^2E_2 =0\\
    \ii \partial_Z E_3 &+ \partial_Y E_2 +\ii \nu_0 E_1 + \widetilde{\sigma} \left|E_3\right|^2 E_3 =0.
\end{align}
    \label{Eq: NLLieb}
\end{subequations}
They appear to be %are 
a novel nonlinear system. The corresponding dispersion relations are
\begin{equation}
\mu(K,L) = 0, ~~ \pm \sqrt{\nu_0^2 + K^2 + L^2} .
\end{equation}
with a gap width of $2|\nu_0|$ at the origin. Comparing these continuous eigenvalues to the discrete versions in equation (\ref{lieb_spec}), we observe a natural extension including a zero plane separating two hyperboloid sheets. Moreover, due to the vanishing of one of the dispersion relation functions localized initial values will generally not decay for large $z$.

\begin{figure}[ht]
    \centering
            \includegraphics[width=0.8\textwidth]{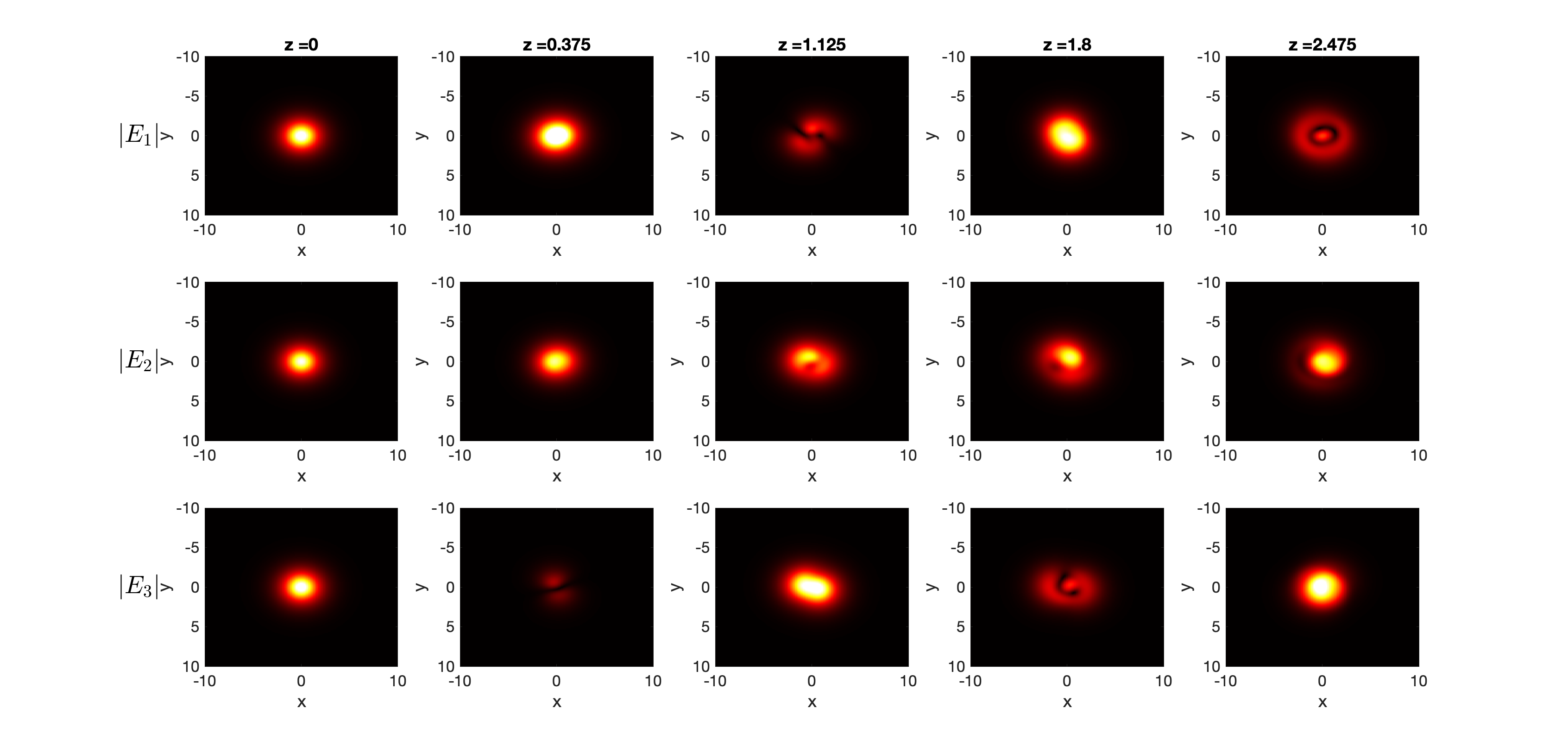}
            \caption{Evolution of an initial pulse in the nonlinear Lieb system (\ref{Eq: NLLieb}) with $\nu_0 = 2$ and $\widetilde{\sigma}= 0.6$. Energy oscillates between the $E_1$ and $E_3$ modes, mediated by $E_2$.}
            \label{Fig: NLLieb}
\end{figure}

Again we consider initial conditions where a Gaussian envelope is taken on all (three) eigenfunctions. In Figure \ref{Fig: NLLieb}, we see an example nonlinear evolution for a pulse in which the energy of the system oscillates between the first and third components of the system. These oscillations are a feature of the related linear system and the focusing nonlinearity effect. 
Similar to the nonlinear Dirac system, the nonlinear Lieb system transitions to a regime of wave collapse as the nonlinear coefficient is increased.

\section{Conclusions}

Tight-binding approximations are constructed for a longitudinally driven two-dimensional lattices with uniform lattice sites. 
The general lattice is doubly periodic in the transverse directions and consists of an arbitrary number of sublattice sites within each unit cell. 
An orbital expansion results in a system of ODE's with order equal to the number of sublattice sites. For fast oscillations in the longitudinal direction, an averaged equation is derived with constant coefficients in terms of $z$ and parameterized by the transverse wavenumber, $\bm k$. 

These averaged models have the effect of introducing next-nearest neighbor interactions into the system. For the honeycomb lattice the averaging process leads to the well known Haldane Model in Fourier space.
Analytic calculation of the Berry connection and curvature means that global information about the topology of the spectral bands can be obtained. Only a local approximation of the eigenvectors is required for the connection and curvature since the Chern number is determined by the local behavior near discontinuities. Detailed analysis is also carried out for the Lieb lattice.

Nonlinear wave equations are derived that govern the envelopes of the underlying eigenvectors. In the case of a honeycomb lattice  the resulting nonlinear Dirac system contains an extra "mass" term, in comparison to previous models \cite{Nixon2009, Zhu2010}. Numerical computation  shows spiral diffraction patterns for relatively small nonlinear effects. As the effect of the nonlinearity is increased the wave envelopes tend towards a soliton type structure and at larger nonlinear coefficent the envelopes undergo collapse. For Dirac points of the spectrum with a flatband all crossing the point, as appears in the Lieb and $1/5$-depleted lattices, a novel nonlinear system is derived for the envelope dynamics. In this system simulations show energy oscillates between the eigenfunctions.

The authors have no competing interests to declare.

This article has no additional data.

\section*{Acknowledge}
This work was partially supported by AFOSR under Grant
No. FA9550-19-1-0084 and NSF under Grant
No. DMS-2005343.

\bibliographystyle{unsrtnat}
\bibliography{ChernBib.bib}

\appendix                                     
\section{Tight binding details}
\label{APP: Details}

A heuristic understanding of the $\mathcal{L}$ operators in (\ref{Eq: RealSystem}) can be reached by considering a lattice with a single pair of lattice sites centered at $\rmvec{r}_1$ and $\rmvec{r}_2$, i.e., $V(\rmvec{r}) = V_0^2 - \widetilde{V}(\rmvec{r}-\rmvec{r}_1) - \widetilde{V}(\rmvec{r}-\rmvec{r}_2)$. This corresponds to an approximation with scalar coefficient functions $a_1(z)$ and $a_2(z)$, 
\begin{equation}
    \psi(\rmvec{r},z) \approx \Big( a_1(z) \phi_1(\rmvec{r})
     + a_2(z) \phi_2(\rmvec{r})\Big) \ee^{-\ii \left(\mu +V_0^2\right)z},
\end{equation}
where $\phi_j(\rmvec{r})= \phi_0(\rmvec{r}-\rmvec{r}_j)$. Substituting this directly into equation (\ref{Eq: PSEMovingFrame}) gives  
\begin{align}
\label{Eq: TwoLatticePlugIn}
    0=&~~\ii a_1'(z) \phi_1 - \ii a_1(z) \nabla \phi_1 \cdot \rmvec{h}'(z)
        + a_1(z) \widetilde{V}_2 \phi_1  \\
   &+ \ii a_2'(z) \phi_1 - \ii a_2(z) \nabla \phi_2 \cdot \rmvec{h}'(z) 
        + a_2(z) \widetilde{V}_1 \phi_2 \notag\\  
        &+ \sigma \Big( \lvert a_1 \rvert^2 a_1 \phi_1^3 + 2 \lvert a_1 \rvert^2 a_2 \phi_1^2\phi_2 +a_1^2 a_2^* \phi_1^2\phi_2\notag\\
        &~~~~~~+\lvert a_1 \rvert^2 a_2 \phi_2^3 + 2 \lvert a_2 \rvert^2 a_1 \phi_2^2\phi_1 +a_2^2 a_1^* \phi_2^2\phi_1\Big)\notag
\end{align}
where $\widetilde{V}_j(\rmvec{r})= \widetilde{V}(\rmvec{r}-\rmvec{r}_j)$. Taking the inner product of (\ref{Eq: TwoLatticePlugIn}) with $\phi_j$, we keep terms that scale with $V_0$ and are large due to the deep lattice assumption, $V_0\gg 1$. This includes terms with $\widetilde{V}$ and terms with $\nabla\phi_j$ that scale with $V_0$ due to the rapid decay rate of the tails. Otherwise, tail interactions are exponentially small and we drop all but the onsite nonlinear effects. This results in the system
\begin{subequations}
\begin{align}
    \ii \frac{\dd a_1}{\dd z} & + p_0 a_1 
     + \Big(q_0 - \ii s_0 \rmvec{h}'(z)\cdot(\rmvec{r}_2 -\rmvec{r}_1)\Big) a_2+ \widetilde{\sigma}\left|a_1\right|^2a_1=0\\
      \ii \frac{\dd a_2}{\dd z} & + p_0 a_2 
      + \Big(q_0 - \ii s_0 \rmvec{h}'(z)\cdot(\rmvec{r}_1 -\rmvec{r}_2)\Big) a_1+ \widetilde{\sigma}\left|a_2\right|^2a_2=0
\end{align}
\end{subequations}
where all coefficients are determined by the orbital form $\phi_0(\bm r)$ and distance between the nearest neighbor lattice sites. 
\begin{align*}
    &p_0 = \iint \widetilde{V}(\rmvec{r} - \Delta \rmvec{r}) \phi_0(\rmvec{r})^2 \dd r_x \dd r_y,
    \quad  
    q_0 = \iint \widetilde{V} (\rmvec{r})\phi_0(\rmvec{r} - \Delta \rmvec{r})\phi_0(\rmvec{r}) \dd r_x \dd r_y \\
&~~~~~~~~~~~~~~~s_0 \Delta\rmvec{r} \approx \iint \nabla\phi_0(\rmvec{r}- \Delta \rmvec{r})\phi_0(\rmvec{r} ) \dd r_x \dd r_y 
\end{align*}
In the case of the last integral, this is an asymptotic approximation that has been verified against numerical results. This suggests an integrating factor $a_j = \widetilde{a}_j e^{\ii \, p_0 z }$, to remove onsite interactions from the system. 
In term of our binary lattice example this gives 
\begin{subequations}
\begin{align}
    \ii \frac{\dd \widetilde{a}_1}{\dd z}  + \Big(q_0 - \ii s_0 \rmvec{h}'(z)\cdot (\rmvec{r}_2 -\rmvec{r}_1)\Big) \widetilde{a}_2 + \widetilde{\sigma}\left|\widetilde{a}_1\right|^2\widetilde{a}_1&=0\\
      \ii \frac{\dd \widetilde{a}_2}{\dd z} + \Big(q_0 - \ii s_0 \rmvec{h}'(z)\cdot(\rmvec{r}_1 -\rmvec{r}_2)\Big) \widetilde{a}_1+ \widetilde{\sigma}\left|\widetilde{a}_2\right|^2\widetilde{a}_2&=0 .
\end{align}
\end{subequations}
We now see that for $\widetilde{a}_1$ the interaction depends on the vector $\rmvec{r}_2- \rmvec{r}_1$, which points to the neighboring lattice site; and vice versa for $\widetilde{a}_2$. We define the coefficient function given earlier in equation (\ref{Eq: qz}),  
\begin{equation}
    q(z,\rmvec{w}) = q_0 - \ii s_0 \rmvec{h}'(z)\cdot \rmvec{w},
\end{equation}
and the system of equations for a full lattice of the form (\ref{Eq: LatticeFull}) can now be seen as a superposition of all pairwise interactions between nearest neighbor lattice sites.

\end{document}